\documentclass[graybox,natbib]{svmult}

\usepackage{mathptmx}       % selects Times Roman as basic font
\usepackage{helvet}         % selects Helvetica as sans-serif font
\usepackage{courier}        % selects Courier as typewriter font
\usepackage{type1cm}        % activate if the above 3 fonts are
\usepackage{makeidx}         % allows index generation
\usepackage{graphicx}        % standard LaTeX graphics tool
\usepackage{multicol}        % used for the two-column index
\usepackage[bottom]{footmisc}% places footnotes at page bottom

\makeindex             % used for the subject index
\begin{document}

\title*{The First Stars}
% Use \titlerunning{Short Title} for an abbreviated version of
% your contribution title if the original one is too long
\author{Simon Glover}
% Use \authorrunning{Short Title} for an abbreviated version of
% your contribution title if the original one is too long
\institute{Simon Glover \at Universit\"at Heidelberg, Zentrum f\"ur Astronomie, Institut f\"ur Theoretische
Astrophysik, Albert-Ueberle-Str.\ 2, 69120 Heidelberg, \email{glover@uni-heidelberg.de}}
%
% Use the package "url.sty" to avoid
% problems with special characters
% used in your e-mail or web address
%
\maketitle

\abstract{The first stars to form in the Universe -- the so-called Population III stars --
bring an end to the cosmological Dark Ages, and exert an important influence on the 
formation of subsequent generations of stars and on the assembly of the first galaxies.
Developing an understanding of how and when the first Population III stars formed and 
what their properties were is an important goal of modern astrophysical research. In this 
review, I discuss our current understanding of the physical processes involved in the
formation of Population III stars. I show how we can identify the mass scale of the
first dark matter halos to host Population III star formation, and discuss how gas 
undergoes gravitational collapse within these halos, eventually reaching protostellar
densities. I highlight some of the most important physical processes occurring during
this collapse, and indicate the areas where our current understanding remains 
incomplete. Finally, I discuss in some detail the behaviour of the gas after the formation 
of the first Population III protostar. I discuss both the conventional picture, where the gas 
does not undergo further fragmentation and the final stellar mass is set by the interplay 
between protostellar accretion and protostellar feedback, and also the recently advanced
picture in which the gas does fragment and where dynamical interactions between 
fragments have an important influence on the final distribution of stellar masses.}

\section{Formation of the first star-forming minihalo}
\subsection{The Jeans mass and the filter mass}
In the currently dominant $\Lambda$CDM paradigm, gravitationally \index{minihalo|(}
bound objects form in a hierarchical fashion, with the smallest, least
massive objects forming first, and larger objects forming later through 
a mixture of mergers and accretion. The mass scale of the least massive
objects to form from dark matter is set by free-streaming of the dark matter
particles, and so depends on the nature of these particles. However, in
most models, this minimum mass is many orders of magnitude smaller 
than the mass of even the smallest dwarf galaxies \citep{ghs05}. More 
relevant for the formation of the first stars and galaxies is the mass 
scale of the structures (frequently referred to as dark matter `minihalos')
within which the baryonic component of matter, the gas, can first cool and 
collapse.

A lower limit on this mass scale comes from the theory of the growth of
small density perturbations in an expanding universe \citep[see e.g.][]{bl01}.
From the analysis of the linearized equations of motion, one can identify a critical 
length scale, termed the Jeans length, that marks the boundary between gravitationally   \index{Jeans length}
stable and gravitationally unstable regimes. The Jeans length is given (in physical units) by
\begin{equation}
\lambda_{\rm J} = c_{\rm s} \sqrt{\frac{\pi}{G \rho_{0}}},
\end{equation}
where $c_{\rm s}$ is the sound speed in the unperturbed intergalactic medium and
$\rho_{0}$ is the cosmological background density. Perturbations on scales 
$\lambda > \lambda_{\rm J}$ are able to grow under the influence of
their own self-gravity, while those with $\lambda < \lambda_{\rm J}$ are
prevented from growing by thermal pressure. We can associate a mass
scale with $\lambda_{\rm J}$ by simply taking the mass within a sphere
of radius $\lambda_{\rm J} / 2$ \citep{bl01}:
\begin{equation}
M_{\rm J} = \frac{4\pi}{3} \rho_{0} \left(\frac{\lambda_{\rm J}}{2}\right)^3.
\end{equation}
This mass, termed the Jeans mass, describes the minimum mass that a perturbation  \index{Jeans mass|(}
must have in order to be gravitationally unstable.

\begin{figure}
\includegraphics[scale=.45]{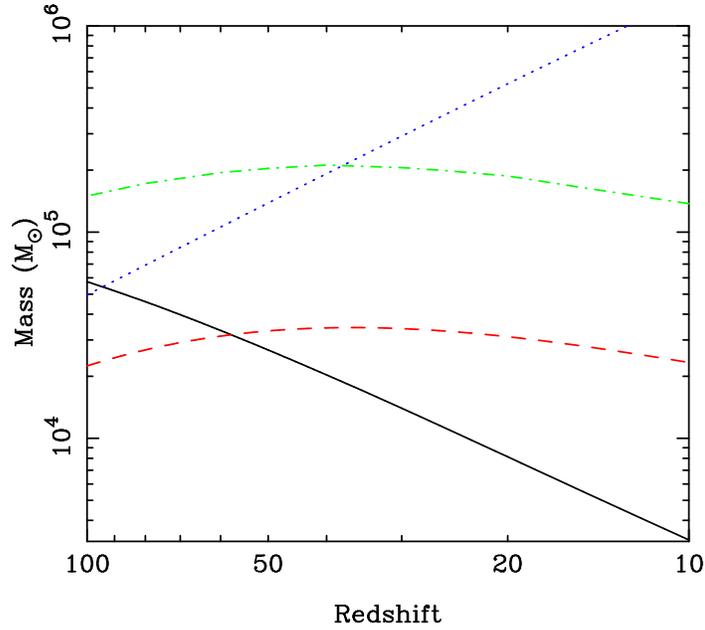}
\caption{Evolution with redshift of the Jeans mass (solid line), the filter mass computed in the
limit where the relative streaming velocity between gas and dark matter is zero (dashed line) and 
the filter mass computed assuming a streaming velocity $v = \sigma_{\rm vbc}$ 
(dash-dotted line). Also plotted is the critical minihalo mass, $M_{\rm crit}$, required for efficient 
H$_{2}$ cooling (dotted line). The estimate of the filter mass in the no streaming limit comes from
\citet{nb07}, who account for a number of effects not treated in the original \citet{gh98} formulation,
while the estimate of $M_{\rm F}$ in the streaming case comes from \citet{tbh10}. The value of
$M_{\rm crit}$ was computed using Equation~\ref{mcrit}.}
\label{fig:mass} 
\end{figure}

In the simplest version of this analysis, the value used for the sound speed in
the equations for the Jeans length and Jeans mass is the instantaneous value;
i.e.\ to determine $\lambda_{\rm J}$ and $M_{\rm J}$ at a redshift $z$, we use the
value of $c_{s}$ at that redshift. In this approximation, the Jeans mass is given in the 
high redshift limit (where the gas temperature is strongly coupled to the cosmic microwave 
background [CMB] temperature by Compton scattering) by the expression \citep{bl01}
\begin{equation}
M_{\rm J} = 1.35 \times 10^{5} \left(\frac{\Omega_{m} h^{2}}{0.15} \right)^{-1/2} \: {\rm M_{\odot}},
\end{equation}
where $\Omega_{\rm m}$ is the dimensionless cosmological matter density parameter, and 
$h$ is the value of the Hubble constant in units of $100 \: {\rm km} \: {\rm s^{-1}} \: {\rm Mpc^{-1}}$.
In the low redshift limit (where the coupling between radiation and matter is weak and
the gas temperature evolves adiabatically), the Jeans mass is given instead by
\begin{equation}
M_{\rm J} = 5.18 \times 10^{3} \left(\frac{\Omega_{m} h^{2}}{0.15} \right)^{-1/2} \left(
\frac{\Omega_{b} h^{2}}{0.026}\right)^{-3/5} \left(\frac{1+z}{10}\right)^{3/2} \: {\rm M_{\odot}},
\end{equation}
where $\Omega_{\rm b}$ is the dimensionless cosmological baryon density parameter.
The evolution of $M_{\rm J}$ with redshift is also illustrated in Figure~\ref{fig:mass}. \index{Jeans mass|)}

A more careful treatment of the growth of linear density perturbations accounts for the fact
that the sound speed, the Jeans length and potentially also the Jeans mass may all change \index{Jeans length} \index{Jeans mass}
significantly during the time it takes for a perturbation to grow into the non-linear regime.
\citet{gh98} showed that in this case, the appropriate mass scale separating the gravitationally
stable and gravitationally unstable regimes is a form of time-averaged Jeans mass that they
denote as the ``filter mass'',  $M_{\rm F}$. This is given in physical units by  \index{filter mass|(}
\begin{equation}
M_{\rm F} = \frac{4\pi}{3} \rho_{0} \left(\frac{\lambda_{\rm F}}{2}\right)^3,  \label{mf}
\end{equation}
where the filter wavelength $\lambda_{\rm F}$ is given in the high redshift limit by
\citep{gn00}
\begin{equation}
\lambda_{\rm F}^{2} = \frac{3}{1+z} \int_{z}^{\infty} \lambda_{\rm J}^{2} 
\left[1 - \left(\frac{1+z}{1+z^{\prime}} \right)^{1/2} \right] \, {\rm d}z^{\prime}.
\label{lf}
\end{equation}

It is possible to improve further on this analysis by accounting for spatial variations in
the sound speed \citep{bl05,nb05} and by properly accounting for the separate rates
of growth of the dark matter and baryonic perturbations in the high redshift limit in 
which the gas is mechanically coupled to the CMB by Compton scattering 
\citep{nb07}. The net result is to somewhat lower the filter mass in comparison with 
the predictions of Equations~\ref{mf}-\ref{lf}. Comparing the resulting filter mass with
the Jeans mass (Figure 1), we see that the filter mass can be a factor of a few
smaller than the Jeans mass at high redshift, but that for redshifts below $z \sim 50$,  \index{Jeans mass}
the filter mass is the larger of the two mass scales.

Another complication was recently pointed out by \citet{th10}. They show that prior  \index{streaming velocity|(}
to the recombination epoch, the strong coupling between gas and radiation leads
to the gas developing a non-zero velocity relative to the dark matter. While the
gas and radiation are coupled, the sound-speed in the gas is approximately
$c / \sqrt{3}$, where $c$ is the speed of light, and the relative velocity between gas
and dark matter is highly subsonic. Once the gas and radiation decouple, however,
the sound-speed of the gas decreases enormously, becoming $\sim 6 \: {\rm km}
\: {\rm s^{-1}}$ at the end of the recombination epoch. \citet{th10} show that at the same
time, the RMS velocity of the gas relative to the dark matter is about $30 \:
{\rm km} \: {\rm s^{-1}}$, implying that the gas is moving supersonically with respect to
the dark matter. The coherence length of the supersonic flow is of the order of the Silk 
damping scale \citep{silk68}, i.e.\ several comoving Mpc, and so on the much smaller 
scales corresponding to the formation of the first star-forming minihalos, the gas can
be treated as being in uniform motion with respect to the dark matter. \citet{th10} also
show that the relative velocity between gas and dark matter acts to suppress the growth
of small-scale structure in both components, and that because this effect is formally a
second-order term in cosmological perturbation theory, it was not included in previous
studies based on linear perturbation theory.

In a follow-up study, \citet{tbh10} improve on the \citet{th10} analysis by accounting for
spatial variations in the sound speed, and study the effect that the relative velocity between
the gas and the dark matter has on the size of the filter mass. The magnitude of the relative
velocity $v$ is randomly distributed with a Gaussian probability distribution function (PDF) with
total variance $\sigma^{2}_{\rm vbc}$, i.e.
\begin{equation}
P_{\rm vbc}(v) = \left(\frac{3}{2 \pi \sigma_{\rm vbc}^{2}}\right)^{3/2} 4\pi v^{2}
\exp \left(-\frac{3 v^{2}}{2 \sigma^{2}_{\rm vbc}} \right).   \label{vbc_pdf}
\end{equation}
\citet{tbh10} show that for a relative velocity $v = \sigma_{\rm vbc}$ (i.e. a one
sigma perturbation), the effect of the relative velocity between gas and dark matter is
to increase $M_{\rm F}$ by roughly an order of magnitude, as illustrated in 
Figure~\ref{fig:mass}. Higher sigma perturbations lead to even greater increases in 
$M_{\rm F}$, but \citet{tbh10} show that the global average case (obtained by computing 
$M_{\rm F}$ for a range of different $v$ and then integrating over the PDF given in
Equation~\ref{vbc_pdf}) is very similar to the one sigma case.  Numerical studies of the
effects of these streaming velocities \citep[see e.g.][]{sbl11,gr11b} have generally confirmed this result, 
although these studies still disagree somewhat regarding the influence of the streaming
velocities on minihalos with masses greater than the revised $M_{\rm F}$.   \index{filter mass|)} 
\index{streaming velocity|)}

Nevertheless, even the most careful version of this analysis only tells us the mass scale 
of the first gravitationally bound structures to have a significant gas content, which is 
merely a lower limit on the mass scale of the first {\em star-forming} minihalos. The reason 
for this is that for stars to form within a minihalo, it is not enough that the gas be gravitationally 
bound; it must also be able to cool efficiently. In order for the gas within a minihalo to dissipate 
a large fraction of its gravitational binding energy -- a necessary condition if pressure
forces are not to halt the gravitational collapse of the gas \citep{hoyle53,rees76,ro77}
-- it must be able to radiate this energy away. The timescale over which this occurs is
known as the cooling time, and is defined as \index{cooling time}
\begin{equation}
t_{\rm cool} = \frac{1}{\gamma - 1} \frac{n_{\rm tot} k T}{\Lambda},
\end{equation}
where $n_{\rm tot}$ is the total number density of particles, $\gamma$ is the adiabatic
index, $k$ is Boltzmann's constant, $T$ is the gas temperature and 
$\Lambda$ is the radiative cooling rate per unit volume. If the cooling
time of the gas is longer than the Hubble time, then it is very unlikely that the minihalo
will survive as an isolated object for long enough to form stars. Instead, it is far more likely
that it will undergo a major merger with another dark matter halo of comparable or larger
mass before any of its gas has cooled significantly, since major mergers occur, on average,
approximately once per Hubble time \citep{lc93}. Therefore, to determine the minimum
mass of a star-forming minihalo, we must first understand how cooling occurs within 
primordial gas, a topic that we explore in the next section.

\subsection{Cooling and chemistry in primordial gas}
\label{coolchem}
At high temperatures ($T \sim 10^{4} \: {\rm K}$ and above), primordial gas can cool efficiently 
through the collisional excitation of excited electronic states of atomic hydrogen, atomic helium,
and singly-ionized helium. However, it is relatively easy to show that most of the gas within a 
minihalo with $M \sim M_{\rm F}$ will have a temperature significantly below $10^{4} \: {\rm K}$.
If we assume that the gas within the minihalo relaxes into a state of virial equilibrium, such that
the total potential energy $W$ and total kinetic energy $K$ are related by $W = - 2K$, then we
can use this fact to define a virial temperature for the minihalo \citep{bl01}  \index{virial temperature}
\begin{equation}
T_{\rm vir} =  \frac{\mu m_{\rm p} v_{\rm c}^{2}}{2 k},  
\end{equation}
where $\mu$ is the mean molecular weight, $m_{\rm p}$ is the proton mass, and 
$v_{c}$ is the circular velocity of the minihalo. This can be rewritten in terms of the redshift $z$
and the mass $M$ of the minihalo as
\begin{equation}
T_{\rm vir} = 1.98 \times 10^{4} \left(\frac{\mu}{0.6}\right) \left(\frac{M}{10^{8} h^{-1} \: {\rm M_{\odot}}}
\right)^{2/3} \left[\frac{\Omega_{m}}{\Omega_{m}(z)} \frac{\Delta_{c}}{18\pi^{2}} \right]^{1/3} \left(
\frac{1+z}{10} \right) \: {\rm K},  \label{tvir}
\end{equation}
where $\Omega_{m}(z)$ is the dimensionless cosmological density parameter evaluated at 
redshift $z$ and $\Delta_{c} = 18 \pi^{2} + 82 d - 39 d^{2}$, with $d = \Omega_{m}(z) - 1$
\citep{bn98}. In the standard $\Lambda$CDM cosmology, $\Omega_{m}(z) \simeq 1$ at
$z > 6$, and hence the term in square brackets reduces to $\Omega_{m}^{1/3}$. If we rearrange 
Equation~\ref{tvir}  and solve for the mass $M_{\rm atom}$ of a cloud that has a virial temperature 
$T_{\rm vir} = 10^{4} \: {\rm K}$ and that can therefore cool via atomic excitation, we find that 
\begin{equation}
M_{\rm atom} =  5 \times 10^{7} h^{-1} \left(\frac{\mu}{0.6}\right)^{-3/2} \Omega_{m}^{-1/2} 
\left(\frac{1+z}{10} \right)^{-3/2}  \: {\rm M_{\odot}},
\end{equation}
significantly larger than our estimates for $M_{\rm J}$ and $M_{\rm F}$ above. Minihalos with masses 
close to $M_{\rm J}$ or $M_{\rm F}$ will therefore have virial temperatures much less than $10^{4} \: {\rm K}$,
placing them in the regime where molecular coolants dominate.

In primordial gas, by far the most abundant and hence most important molecule is molecular 
hydrogen,  \index{molecular hydrogen!chemistry|(}
H$_{2}$. The chemistry of H$_{2}$ in primordial gas has been reviewed in a number of different studies
\citep[see e.g.][]{aazn97,gp98,sld98,ga08}, and so we only briefly discuss it here. Direct formation of H$_{2}$ by 
the radiative association of two hydrogen atoms is highly forbidden \citep{gs63}, and so at low
densities, most H$_{2}$ forms via the reaction chain \citep{md61,pd68}
\begin{eqnarray}
{\rm H} + {\rm e^{-}} & \rightarrow & {\rm H^{-}} + \gamma, \label{ra} \\
{\rm H^{-}} + {\rm H} & \rightarrow & {\rm H_{2}} + {\rm e^{-}},  \label{ad}
\end{eqnarray}
with a minor fraction forming via the reaction chain \citep{sz67}
\begin{eqnarray}
{\rm H} + {\rm H^{+}} & \rightarrow & {\rm H_{2}^{+}} + \gamma, \\
{\rm H_{2}^{+}} + {\rm H} & \rightarrow & {\rm H_{2}} + {\rm H^{+}}.
\end{eqnarray}
In warm gas, H$_{2}$ can be destroyed by collisional dissociation
\citep[see e.g.][]{mkm98}
\begin{eqnarray}
{\rm H_{2}} + {\rm H} & \rightarrow & {\rm H} + {\rm H} + {\rm H}, \\
 {\rm H_{2}} + {\rm H_{2}} & \rightarrow & {\rm H} + {\rm H} + {\rm H_{2}},
\end{eqnarray}
or by charge transfer with H$^{+}$ \citep[see e.g.][]{savin04}
\begin{equation}
{\rm H_{2}} + {\rm H^{+}} \rightarrow {\rm H_{2}^{+}} + {\rm H},
\end{equation}
but at low temperatures there are no collisional processes that can efficiently
remove it from the gas.

When the fractional ionization of the gas is low, the rate at which H$_{2}$
forms is limited primarily by the rate at which H$^{-}$ ions form via 
reaction~\ref{ra}, as any ions that form are rapidly converted to H$_{2}$ by
associative detachment with atomic hydrogen (reaction~\ref{ad}). 
If the fractional ionization is large, on the other hand, then
many of the H$^{-}$ ions formed by reaction~\ref{ra} do not survive for long
enough to form H$_{2}$, but instead are destroyed by mutual neutralization
with H$^{+}$ ions:
\begin{equation}
{\rm H^{+}} + {\rm H^{-}} \rightarrow {\rm H + H}. \label{mn}
\end{equation}
The ratio of the rates of reactions~\ref{ad} and \ref{mn} is given by
$k_{\ref{ad}} n_{\rm H} / k_{\ref{mn}} n_{\rm H^{+}}$, where $n_{\rm H}$
is the number density of atomic hydrogen, $n_{\rm H^{+}}$ is the number
density of protons, and $k_{\ref{ad}}$ and $k_{\ref{mn}}$ are the rate
coefficients for reactions \ref{ad} and \ref{mn}, respectively. Mutual neutralization
therefore becomes significant whenever $n_{\rm H^{+}} / n_{\rm H}
\geq k_{\ref{ad}} / k_{\ref{mn}}$. Although the value of $k_{\ref{ad}} / k_{\ref{mn}}$
is temperature dependent, the temperature dependence is weak if one uses
the best available determinations of the rate coefficients (\citealt{kreck10} 
for reaction~\ref{ad}, \citealt{stenrup09} for reaction~\ref{mn}\footnote{A group
lead by X.~Urbain at the Universit\'{e} Catholique de Louvain 
has recently made new experimental measurements
of the rate of this reaction at low temperatures, but at the time of writing, the
results of this work remain unpublished}), and   
$k_{\ref{ad}} / k_{\ref{mn}} \sim 0.03$ to within 50\% for all temperatures
$100 < T < 10^{4} \: {\rm K}$. If we compare this value with the residual
fractional ionization of the intergalactic medium (IGM) at this epoch, 
$x \sim 2 \times 10^{-4}$ \citep{sch08},
we see that mutual neutralization is unimportant within the very first star-forming
minihalos. It becomes important once larger minihalos, with virial temperatures
$T_{\rm vir} \sim 10^{4} \: {\rm K}$ or above, begin to form, as in these minihalos,
substantial collisional ionization of the gas can occur, leading to an initial
fractional ionization much higher than the residual value in the IGM. It also
becomes an important process within the ``fossil'' HII regions left behind by the  \index{fossil HII region}
first generation of massive stars \citep{oh03,no05,gsj06,kreck10}.  \index{molecular hydrogen!chemistry|)}

Although H$_{2}$ is by far the most abundant primordial molecule, it is actually  \index{molecular hydrogen!cooling|(}
not a particularly efficient coolant. The H$_{2}$ molecule has no dipole moment,
and so dipole transitions between its excited rotational and vibrational levels are
forbidden. Although radiative transitions between levels do occur, they are 
quadrupole transitions and the associated transition rates are small. In addition,
application of the Pauli exclusion principle to the hydrogen molecule shows that
it must have two distinct states, distinguished by the nuclear spin of the two hydrogen
nuclei: para-hydrogen, in which the nuclear spins are parallel, and which must have
an even value for the rotational quantum number $J$, and ortho-hydrogen, which has
anti-parallel nuclear spins and an odd value for $J$. Radiative transitions between
ortho-hydrogen and para-hydrogen involve a change in orientation of the spin of one
of the nuclei and are therefore strongly forbidden. As a result, the least energetic
rotational transition of H$_{2}$ that has any significant probability of occurring is
the transition between the $J = 2$ and $J = 0$ rotational levels in the vibrational
ground-state of para-hydrogen. This transition has an associated energy 
$E_{20} / k \simeq 512$~K. The H$_{2}$ molecule therefore has large energy
separations between the ground state and any of the accessible excited 
rotational or vibrational states\footnote{For comparison, note that the energy
separation between the $J = 0$ and $J = 1$ rotational levels of CO is roughly 5~K.}, 
and has only weak radiative transitions between these states. 

These features of the H$_{2}$ molecule have two important consequences. First,
it becomes a very inefficient coolant at temperatures $T \ll E_{20} / K$, as it becomes
almost impossible to collisionally populate any of the excited states. The minimum
temperature that can be reached solely with H$_{2}$ cooling depends somewhat 
on the H$_2$ abundance and the time available for cooling, but typically 
$T_{\rm min} \sim 150$-200~K. Second, its rotational and vibrational
levels reach their local thermodynamic equilibrium (LTE) level populations at a
relatively low density, $n_{\rm crit} \sim 10^{4} \: {\rm cm^{-3}}$. This
means that at densities $n \gg n_{\rm crit}$, the H$_{2}$ cooling rate scales only
linearly with density and the cooling time due to H$_2$ becomes independent  \index{cooling time}
of density. Since other important timescales, such as the free-fall collapse time
of the gas, continue to decrease with increasing density, the implication is that
H$_{2}$ becomes an increasingly ineffective coolant as one moves to higher
densities.  \index{molecular hydrogen!cooling|)}

For these reasons, primordial molecules or molecular ions that do not share
these drawbacks have attracted a certain amount of attention. In an early
study, \citet{ls84} suggested that deuterated hydrogen, HD, and lithium hydride,
LiH, may both be significant coolants in primordial gas. More recently, work by
\citet{yokh07} has suggested that H$_{2}^{+}$ may be an important coolant
in some circumstances, while \citet{gs06} show that H$_{3}^{+}$ is also worthy
of attention. In practice, the only one of these molecules or ions that has proved
to be important is HD. Detailed modelling of the chemistry of lithium in primordial 
gas \citep[e.g.][]{sld96,mon05} has shown that LiH is efficiently destroyed by the reaction
\begin{equation}
{\rm LiH} + {\rm H} \rightarrow {\rm Li} + {\rm H_{2}},
\end{equation}
and that only a small fraction of the available lithium (which itself has an abundance
of only $5 \times 10^{-10}$ relative to hydrogen; see \citealt{cyb08}) is ever incorporated into LiH. 
Cooling from the molecular ion H$_{2}^{+}$ was re-examined by \citet{gs09}, who showed that 
the collisional excitation rates cited by \citet{gp98} and used as a basis for the fits given
in \citet{yokh07} were a factor of ten too large, and that if the correct rates are used, H$_{2}^{+}$ 
cooling is no longer  important. Finally, \citet{gs09} also examined the possible role played by  
H$_{3}^{+}$ cooling in considerable detail, but found that even if one makes optimistic assumptions
regarding its formation rate and collisional excitation rate, it still contributes to
the total cooling rate at the level of only a few percent, and hence at best is a minor
correction term.

These studies leave HD as the only viable alternative to H$_{2}$ as a coolant of  \index{HD!chemistry|(}
primordial gas. HD has a small, but non-zero dipole moment, giving it radiative \index{HD!cooling|(}
transition rates that are somewhat larger than those of H$_{2}$, resulting in 
a critical density $n_{\rm crit} \sim 10^{6} \: {\rm cm^{-3}}$. Unlike H$_{2}$, 
it is not separated into ortho and para states, and so the lowest energy transition
accessible from the ground state is the $J = 1$ to $J = 0$ rotational transition, 
with an energy $E_{10} / k = 128$~K. Although the cosmological ratio of deuterium
to hydrogen is small [${\rm D/H} = (2.49 \pm 0.17) \times 10^{-5}$; \citealt{cyb08}], 
the ratio of HD to H$_{2}$ can be significantly
boosted in low temperature gas by chemical fractionation. The reaction
\begin{equation}
{\rm H_{2} + D^{+}} \rightarrow {\rm HD + H^{+}}
\end{equation}
that converts H$_{2}$ into HD is exothermic and so proceeds rapidly at all temperatures,
while the inverse reaction
\begin{equation}
{\rm HD + H^{+}} \rightarrow {\rm H_{2} + D^{+}}
\end{equation}
is endothermic and so proceeds very slowly at low temperatures. In equilibrium, these
two reactions produce an HD-to-H$_{2}$ ratio given by 
\begin{equation}
\frac{x_{\rm HD}}{x_{\rm H_{2}}} = 2 \exp \left(\frac{464}{T}\right) [{\rm D/H}],
\end{equation}
where $[{\rm D/H}]$ is the cosmological D:H ratio. Together, these factors render HD a much more
effective coolant than H$_{2}$ in low temperature gas. \index{HD!chemistry|)}

In practice, for HD cooling to take over from H$_{2}$ cooling, the gas must already be
fairly cold, with $T \sim 150$~K  \citep{g08}, and temperatures this low are typically not
reached during the collapse of the first star-forming minihalos, meaning that HD remains
a minor coolant \citep{bcl02}. However, there are a number of situations, typically involving
gas with an enhanced fractional ionization, in which HD cooling does become significant
\citep[see e.g.][]{nu02,no05,jb06,yokh07,mb08,gr08,kreck10}.   \index{HD!cooling|)}

\subsection{The minimum mass scale for collapse}
The relative simplicity of the chemistry discussed in the previous section allows one \index{molecular hydrogen!chemistry|(}
to construct a very simple model that captures the main features of the evolution of the 
H$_{2}$ fraction within low density gas falling into a dark matter minihalo.
We start by assuming that radiative recombination is the only process affecting
the electron abundance, and writing the rate of change of the electron number density as
\begin{equation}
\frac{{\rm d}n_{\rm e}}{{\rm d}t} = - k_{\rm rec} n_{\rm e} n_{\rm H^{+}},
\end{equation}
where $n_{\rm e}$ is the number density of electrons, $n_{\rm H^{+}}$ is the number
density of protons, and $k_{\rm rec}$ is the recombination coefficient. If we 
assume that ionized hydrogen is the only source of free electrons, implying that
$n_{\rm e} = n_{\rm H^{+}}$, and that the temperature remains roughly constant during
the evolution of the gas, then we can solve for the time evolution of the electron
fraction:
\begin{equation}
x = \frac{x_{0}}{1 + k_{\rm rec} n t x_{0}},
\end{equation}
where $x \equiv n_{\rm e} / n$, $n$ is the number density of hydrogen nuclei, and
$x_{0}$ is the initial value of $x$. We next assume that all of the H$_{2}$ forms via
the H$^{-}$ pathway, and that mutual neutralization of H$^{-}$ with H$^{+}$
(reaction~\ref{mn}) is the
only process competing with associative detachment (reaction~\ref{ad}) for the H$^{-}$
ions. In this case, we can write the time evolution of the H$_{2}$ fraction, 
$x_{\rm H_{2}} \equiv n_{\rm H_{2}} / n$, as
\begin{equation}
\frac{{\rm d}x_{\rm H_{2}}}{{\rm d}t} =  k_{\ref{ra}} x n_{\rm H}  \, p_{\rm AD}, \label{xh2}
\end{equation}
where $k_{\ref{ra}}$ is the rate coefficient of reaction~\ref{ra}, the formation of H$^{-}$
by the radiative association of H and e$^{-}$, and $p_{\rm AD}$ is the probability that
any given H$^{-}$ ion will be destroyed by associative detachment rather than by 
mutual neutralization. Given our assumptions above, this can be written as
\begin{equation}
p_{\rm AD} = \frac{k_{\ref{ad}}n_{\rm H}}{k_{\ref{ad}} n_{\rm H} + k_{\ref{mn}} n_{\rm H^{+}}},
\end{equation}
where $k_{\ref{ad}}$ is the rate coefficient for reaction~\ref{ad} and $k_{\ref{mn}}$
is the rate coefficient for reaction~\ref{mn}. If we again assume that $n_{\rm e} =
n_{\rm H^{+}}$, and in addition assume that $n_{\rm H} \simeq n$, then the expression
for $p_{\rm AD}$ can be simplified to
\begin{equation}
p_{\rm AD} = \left(1 + \frac{k_{\ref{mn}}}{k_{\ref{ad}}} x \right)^{-1}.
\end{equation}
Substituting this into Equation~\ref{xh2}, we obtain
\begin{equation}
\frac{{\rm d}x_{\rm H_{2}}}{{\rm d}t} = k_{\ref{ra}} x n_{\rm H} \left(1 + \frac{k_{\ref{mn}}}{k_{\ref{ad}}} x \right)^{-1}.
\end{equation}

If the initial fractional ionization $x_{0} \ll k_{\ref{ad}} / k_{\ref{mn}}$, then the term in parentheses 
is of order unity and this equation has the approximate solution
\begin{eqnarray}
x_{\rm H_{2}} & = & \frac{k_{\ref{ra}}}{k_{\rm rec}} \ln \left(1 + k_{\rm rec} n x_{0} t \right), \\
 & = & \frac{k_{\ref{ra}}}{k_{\rm rec}} \ln \left(1 + t / t_{\rm rec} \right),
\end{eqnarray}
where $t_{\rm rec} = 1 / (k_{\rm rec} n x_{0})$ is the recombination time. The growth of the H$_{2}$ fraction is
therefore logarithmic in time, with most of the H$_{2}$ forming within the first few recombination times. In 
the more complicated case in which $x_{0}$ is comparable to or larger than $k_{\ref{ad}} / k_{\ref{mn}}$,
but still significantly less than unity (so that $n_{\rm H} \sim n$), the H$_{2}$ fraction is given instead by
\begin{equation}
x_{\rm H_{2}}  = \frac{k_{\ref{ra}}}{k_{\rm rec}} \ln \left(\frac{1 +  x_{0} k_{\ref{mn}} / k_{\ref{ad}} + t / t_{\rm rec}}
 {1 + x_{0} k_{\ref{mn}} / k_{\ref{ad}}} \right). \label{xh2_eq}
\end{equation}

From this analysis, we see that the main factor determining the final H$_{2}$ abundance is the ratio 
$k_{\ref{ra}} / k_{\rm rec}$, since for times of the order of a few recombination times, the logarithmic
term in Equation~\ref{xh2_eq} is of order unity, implying that the 
final H$_{2}$ abundance is at most a factor of a few times 
$k_{\ref{ra}} / k_{\rm rec}$. If we use the simple power-law fits to $k_{\ref{ra}}$ and $k_{\rm rec}$
given by \citet{hutch76}, namely $k_{\ref{ra}} = 1.83 \times 10^{-18} T^{0.8779} \: {\rm cm^{3}} \:
{\rm s^{-1}}$ and $k_{\rm rec} = 1.88 \times 10^{-10} T^{-0.644} \: {\rm cm^{3}} \: {\rm s^{-1}}$,
then we can write the ratio of the two rate coefficients as
\begin{equation}
\frac{k_{\ref{ra}}}{k_{\rm rec}} \simeq 10^{-8} T^{1.5219}.
\end{equation}
The amount of H$_{2}$ produced is a strong function of temperature, but is of the order of a few times
$10^{-3}$ for temperatures of a few thousand Kelvin. 
We see therefore that the formation of H$_{2}$ via H$^{-}$ never results in a gas
dominated by H$_{2}$, as the H$_{2}$ abundance always remains much smaller than the abundance
of atomic hydrogen.  \index{molecular hydrogen!chemistry|)}

Given this simple model for the amount of H$_{2}$ that will form in the gas, the obvious next step is \index{molecular hydrogen!cooling|(}
to compare this to the amount of H$_{2}$ that is required to cool the gas efficiently. In order to determine
the H$_{2}$ fraction necessary to significantly cool gas with a temperature $T$ within some specified
fraction of the Hubble time -- say 20\% of $t_{\rm H}$ -- we can simply equate the two timescales, and
solve for the H$_{2}$ fraction. Using our previous definition of the cooling time, we have   \index{cooling time}
\begin{equation}
\frac{1}{\gamma - 1} \frac{n_{\rm tot} k T}{\Lambda_{0}(T) n_{\rm H_{2}}} = 0.2 t_{\rm H},
\end{equation}
where we have assumed that H$_{2}$ is the dominant coolant and have written the cooling rate 
per unit volume in terms of $\Lambda_{0}$, the cooling rate per H$_{2}$ molecule. Rearranging 
this equation, using the fact that when the H$_{2}$ fraction and the ionization level are low,
$\gamma = 5/3$ and $n_{\rm tot} = (1 + 4 x_{\rm He}) n$, where $x_{\rm He}$ is the fractional
abundance of helium (given by $x_{\rm He} = 0.083$ for primordial gas), we obtain
\begin{equation}
x_{\rm H_{2}, req} = 1.38 \times 10^{-15} \frac{T}{\Lambda_{0}(T)} t_{\rm H}^{-1}.
\end{equation}
In the high-redshift limit where $t_{\rm H} \simeq H_{0}^{-1} \Omega_{m}^{-1/2} (1+z)^{-3/2}$,  
this becomes
\begin{equation}
x_{\rm H_{2}, req} = 5.2 \times 10^{-32} \frac{T}{\Lambda_{0}(T)} \left(\frac{1+z}{10}\right)^{3/2},
\end{equation}
where we have used values for the cosmological parameters taken from \citet{wmap7}.
Collisions of H$_{2}$ with a number of different species contribute to $\Lambda_{0}$, as explored
in \citet{ga08}, but in the earliest minihalos, the dominant contributions come from collisions with
H and He. $\Lambda_{0}$ is therefore given to a good approximation by
\begin{equation}
\Lambda_{0} = \Lambda_{H} n_{\rm H} + \Lambda_{\rm He} n_{\rm He}.
\end{equation}
Simple fits for the values of $\Lambda_{\rm H}$ and $\Lambda_{\rm He} $ as a 
function of temperature can be found in \citet{ga08}. 

\begin{figure}
\includegraphics[scale=.45]{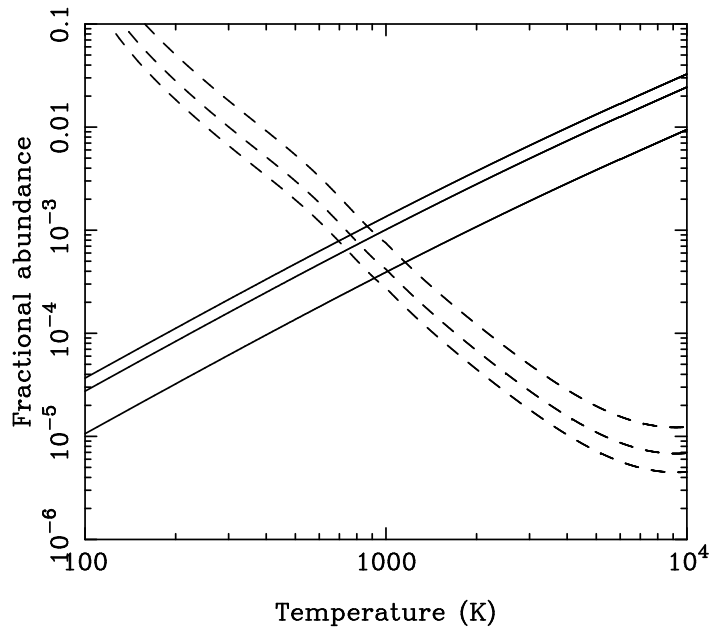}
\caption{Comparison of the fractional abundance of H$_{2}$ produced with our simple toy model for the
chemistry (solid lines) versus the quantity of H$_{2}$ required in order to cool the gas within 20\% of a
Hubble time (dashed lines). From bottom to top, the solid lines correspond to the H$_{2}$ fraction produced
at times $t = 1, 5$ and $10 \: t_{\rm rec}$, respectively, where $t_{\rm rec}$ is the recombination time, and
the dashed lines correspond to the H$_{2}$ fraction required at redshifts $z = 40, 30$ and 20, respectively.
We see that the minimum temperature that the gas must have in order to be able to cool within a fraction of
a Hubble time -- indicated by the point at which the lines cross -- is relatively insensitive to our choices for
$t$ and $z$.
}
\label{fig:h2}  
\end{figure}

An illustration of the  likely size of $x_{\rm H_{2}, req}$  is given in
Figure~\ref{fig:h2}. In this Figure, we plot  $x_{\rm H_{2}, req}$ as a function of temperature, 
evaluated for three different redshifts: $z =20$, 30 and 40. In computing these values, we have assumed 
that the mean density of the gas in the minihalo is given by $\bar{\rho} = \Delta_{c} \rho_{b,0}$, where 
$\rho_{0}$ is the cosmological background density of baryons. In the Figure, we also show the actual H$_{2}$ 
fraction produced in the gas, $x_{\rm H_{2}, act}$, as  a function of temperature at times equal to 1, 5 and 10 
recombination times, and where we have taken  $x_{0} \ll k_{\ref{ad}} / k_{\ref{mn}}$. \index{molecular hydrogen!cooling|)}

Figure~\ref{fig:h2} demonstrates that the amount of H$_{2}$ produced in the gas is a strongly increasing function
of temperature, while the amount required to bring about efficient cooling of the gas is a strongly decreasing
function of temperature. This means that for any given choice of comparison time $t$ and redshift $z$, we can
identify a critical temperature $T_{\rm crit}$, such that gas with $T > T_{\rm crit}$ will cool within a small fraction
of a Hubble time, while gas with $T < T_{\rm crit}$ will not. Moreover, because $x_{\rm H_{2}, act}$ and
$x_{\rm H_{2}, req}$ are both steep functions of temperature, but are relatively insensitive to changes in $t$
or $z$, the value of $T_{\rm crit}$ that we obtain is also relatively insensitive to our choices for $t$ or $z$.  
We find that $T_{\rm crit} \sim 1000$~K, and that at this temperature, the H$_{2}$ fraction required to provide
efficient cooling lies somewhere between a few times $10^{-4}$ and $10^{-3}$ \citep[c.f.][who come to a similar
conclusion using a very similar argument]{teg97}.  If we convert this critical virial temperature into a corresponding
critical minihalo mass using Equation~\ref{tvir}, we find that
\begin{equation}
M_{\rm crit} \simeq 6 \times 10^{5} h^{-1} \left(\frac{\mu}{1.2}\right)^{-3/2} \Omega_{m}^{-1/2} 
\left(\frac{1+z}{10} \right)^{-3/2}  \: {\rm M_{\odot}}. \label{mcrit}
\end{equation}
This mass scale is illustrated by the dotted line in Figure~\ref{fig:mass}. At high redshift, it is smaller than the
filter mass scale corresponding to $v_{\rm bc} = \sigma_{\rm vbc}$, demonstrating that at these redshifts, it
is the streaming of the gas with respect to the dark matter that is the main process limiting the formation of
Population III stars.  Below a redshift of around 40, however, $M_{\rm crit}$ becomes the larger mass scale,
implying that at these lower redshifts, there will be a population of small minihalos that contain a significant
gas fraction, but that do not form stars, because their gas is unable to cool in less than a Hubble time.
These small starless minihalos may be important sinks for ionizing photons during the epoch of reionization 
\citep{ham01}. 

To conclude our discussion of the first star-forming minihalos, we should mention one potentially
important effect not taken into account in the analysis above. This is the influence of ongoing minor mergers
and accretion on the thermal balance of the gas. Although major mergers occur only once per Hubble time,
on average, minor mergers occur far more frequently, and act to stir up the gas, thereby heating it and lengthening
the time required for it to cool. This phenomenon was noted by \citet{yahs03} in their cosmological simulations
of the formation of the first star-forming minihalos. \citet{yahs03} show that in spite of the approximations made
in its derivation, Equation~\ref{mcrit} gives a reasonable guide to the minimum mass of the minihalos that 
contain gas that can cool effectively. However, they also find that there are some minihalos with $M > M_{\rm crit}$
in which the gas does not cool. They show that these minihalos have higher mass accretion rates than
minihalos of the same mass in which cooling does occur, and hence ascribe the suppression of cooling to the
effects of dynamical heating by the ongoing accretion and minor mergers. This effect was also treated more recently 
by \citet{wa08}, who show that it can be included into the simple thermal model described above by the addition
of a heating term describing the effects of mergers and accretion. They show that if one writes this heating term as
\begin{equation}
\Gamma = \frac{k}{\gamma - 1} \frac{{\rm d}T_{\rm vir}}{{\rm d} t},
\end{equation}
then one can relate the rate of change of the virial temperature to the mass growth rate of the minihalo in a
relatively simple fashion. 
\index{minihalo|)}

\section{Gravitational collapse and the formation of the first protostar}
\label{collapse}
\index{protostar!formation|(}
As the analysis in the previous section has shown, gas in minihalos with virial temperatures greater than about
1000~K (corresponding to masses $M \sim 10^{6} \: {\rm M_{\odot}}$) can form enough H$_{2}$ to cool within a small fraction 
of a Hubble time. This reduces the pressure and allows the gas to collapse further under the influence of its own self-gravity. 
As it does so, the value of the Jeans mass decreases. Many early studies of the formation of primordial stars \index{Jeans mass|(}
assumed that as the Jeans mass decreases and the gas becomes more gravitationally unstable, it begins to
undergo hierarchical gravitational fragmentation in a manner similar to that envisaged by \citet{hoyle53},
with the result that at any given moment, the mean fragment mass is approximately equal to the local Jeans mass
\citep[see][for a historical summary of these models]{g05}. In this picture, one could predict the final mass
of the first stars simply by studying the evolution of the Jeans mass. Moreover, since the minimum Jeans mass \index{Jeans mass|)}
reached during the collapse can be estimated with reasonable accuracy on purely thermodynamical grounds
\citep{rees76,llb76}, in this view of Population III star formation, the dynamics of the gas is of secondary importance.
Around ten years ago, however, it first became possible to model the coupled chemical, dynamical and thermal
evolution of the gas within a primordial minihalo using high resolution 3D numerical simulations 
\citep{abn00,abn02,bcl99,bcl02}. These studies showed that the picture outlined above is wrong: the gas does
not undergo hierarchical fragmentation, and so one cannot predict the masses of the first stars simply by studying
the evolution of the Jeans mass. These high resolution numerical simulations, and the many that have followed \index{Jeans mass}
them \citep[e.g.][to name but a few]{yoha06,on07,mb08}, have for the first time given us a clear picture of exactly how 
gravitational collapse proceeds within one of these early minihalos. In the next section, we will discuss the sequence of events 
that occur as we follow the collapse from the minihalo scale all the way down to the scale of a single Population III 
protostar. Following that, in Sections~\ref{dma} and \ref{magn} we discuss two of the main uncertainties remaining in 
our model for the formation of the first Pop.\ III protostar: the role played by heating and ionization arising from dark
matter self-annihilation (Section~\ref{dma}) and the role played by magnetic fields (Section~\ref{magn}).

\subsection{Thermal and chemical evolution of the gas during collapse}

\subsubsection{Initial collapse}
As gas falls into the minihalo from the intergalactic medium, it is shock-heated to a temperature close to $T_{\rm vir}$.
In the post-shock gas, the electron fraction decreases due to radiative recombination, but at the same time H$_{2}$
forms, primarily via reactions~\ref{ra} and \ref{ad}. As we have already seen, the H$_{2}$ fraction evolves logarithmically
with time, with most of the H$_{2}$ forming within the first few recombination times. As the H$_{2}$ fraction increases,
so does its ability to cool the gas, and so the gas temperature slowly decreases, reducing the pressure and allowing
the gas to collapse to the centre of the minihalo. 

At this point, the evolution of the gas depends upon how much H$_{2}$ it has formed. There are two main outcomes,
and which one occurs within a given minihalo depends primarily on the initial ionization state of the gas. 

\paragraph{\bf The low ionization case}
During the formation of the very first Population III stars (also known as Population III.1, to use the \index{molecular hydrogen!cooling|(}
terminology introduced by \citealt{tm08}), the initial fractional ionization of the gas is the same as the residual ionization
in the intergalactic medium, i.e.\ $x_{0} \sim 2 \times 10^{-4}$. In this case, the amount of H$_{2}$ that forms in the gas
is typically enough to cool it to a temperature of $T \sim 200 \: {\rm K}$ but not below. At this temperature, chemical
fractionation has already increased the HD/H$_{2}$ ratio by a factor of 20 compared to the cosmic deuterium-to-hydrogen
ratio, and as a consequence, HD is starting to become an important coolant. However, the amount of HD that forms in the
gas is not enough to cool it significantly below 200~K \citep{bcl02}, and H$_{2}$ continues to dominate the cooling and
control the further evolution of the gas. In this scenario, the collapse of the gas is greatly slowed once its temperature
reaches 200~K and its density reaches a value of around $10^{4} \: {\rm cm^{-3}}$, corresponding to the critical density
$n_{\rm crit}$, at which the rotational and vibrational level populations of H$_{2}$ approach their local thermodynamic equilibrium (LTE) values. At densities higher than this critical density, the H$_{2}$ cooling rate per unit volume scales 
only linearly with $n$ (compared to a quadratic dependence, $\Lambda_{\rm H_{2}} \propto n^{2}$ at lower densities),
while processes such as compressional heating continue to increase more rapidly with $n$. As a result, the gas 
temperature begins to increase once the density exceeds $n_{\rm crit}$. 

Gas reaching this point in the collapse enters what \citet{bcl02} term a ``loitering'' phase, during which cold gas 
accumulates in the centre of the halo but
only slowly increases its density. This loitering phase ends once the mass of cold gas that has accumulated exceeds
the local value of the Bonnor-Ebert mass  \citep{bonnor56,ebert55}, given in this case by \citep{abn02}
\begin{equation}
M_{\rm BE}  \simeq  40 T^{3/2} n^{-1/2} \: {\rm M_{\odot}},
\end{equation}
which for $n \sim 10^{4} \: {\rm cm^{-3}}$ and $T \sim 200$~K yields $M_{\rm BE} \sim 1000 \: {\rm M_{\odot}}$.\footnote{
Discussions of Population III star formation often refer to the cold clump of gas at the centre of the minihalo as a ``fragment'',
and speak of $M_{\rm BE}$ as the ``fragmentation mass scale'', but in the case of the very first generation of 
star-forming minihalos, this is actually something of a misnomer, as very seldom does more than one ``fragment'' form in 
a given minihalo} Once the mass of cold gas exceeds $M_{\rm BE}$, its collapse speeds up again, and becomes largely decoupled 
from the larger-scale behaviour of the gas. The next notable event to occur in the gas is the onset of three-body 
H$_{2}$ formation, which is discussed in the next section. 

\paragraph{\bf The high ionization case}
If the \index{HD!cooling|(}
initial fractional ionization of the gas is significantly higher than the residual fraction in the IGM, then a slightly
different chain of events can occur. A larger initial fractional ionization implies a shorter recombination time, and
hence a logarithmic increase in the amount of H$_{2}$ formed after a given physical time. An increase in the H$_{2}$
fraction allows the gas to cool to a slightly lower temperature, and hence boosts the HD abundance in two ways:
the lower temperature increases the HD/H$_{2}$ ratio produced by fractionation, and the H$_{2}$ fraction itself is
larger, so any given HD/H$_{2}$ ratio corresponds to a higher HD abundance than in the low ionization case. If
the addition ionization allows enough H$_{2}$ to be produced to cool the gas to $T \sim 150 \: {\rm K}$ (which
requires roughly a factor of three more H$_{2}$ than is required to reach 200~K), then chemical fractionation 
increases the HD abundance to such an extent that it takes over as the dominant coolant \citep{g08}.
This allows the gas to cool further, in some cases reaching a temperature as low as the CMB temperature, 
$T_{\rm CMB}$ \citep[e.g.][]{nu02,no05,jb06,yokh07,mb08,kreck10}. The higher critical density of HD,
$n_{\rm crit, HD} \sim 10^{6} \: {\rm cm^{-3}}$, means that the gas does not reach the loitering phase until
much later in its collapse. Once the gas does reach this phase, however, its subsequent evolution is very
similar to that in the low-ionization case discussed above. Cold gas accumulates at $n \sim n_{\rm crit}$
until its mass exceeds the Bonnor-Ebert mass, which in this case is $M_{\rm BE} \sim 40 \: {\rm M_{\odot}}$
if $T = 100$~K and $n = 10^{6} \: {\rm cm^{-3}}$. Once the gas mass exceeds $M_{\rm BE}$, the collapse
speeds up again, and the gas begins to heat up. Aside from the substantial difference in the size of $M_{\rm BE}$,
the main difference between the evolution of the gas in this case and in the low ionization case lies in the fact
that in the high ionization case, the gas reheats from $T \sim 200$~K or below to $T \sim 1000$~K much
more rapidly than in the low ionization case. As we shall see later, this period of rapid heating has a profound
influence on the ability of the gas to fragment.

Several different scenarios have been identified that lead to an enhanced fractional ionization in the gas, and
that potentially allow the gas to reach the HD-dominated regime. Gas within minihalos with $T_{\rm vir} >
9000$~K will become hot enough for collisional ionization of hydrogen to supply the necessary electrons.
However, as halos of this size will typically have at least one star-forming progenitor \citep{jgb08}, it is
questionable whether many Pop.\ III stars will form in such minihalos, as we would expect the gas in 
most of them to have been enriched with metals by one or more previous episodes of star formation.

Another possibility that has attracted significant attention involves the gas in the minihalo being drawn
from a ``fossil'' HII region, i.e.\ a region that was formerly ionized by a previous Population III protostar  \index{fossil HII region}
but has now recombined \citep[see e.g.][]{oh03,no05,yokh07}. Many studies have shown that the volume
of the IGM ionized by a single massive Pop.\ III star is significantly larger than the volume that is enriched
by the metals produced in the supernova occurring at the end of the massive star's life (see e.g.\ the recent
treatment by \citealt{gg10}, or \citet{cf05} for a summary of earlier work). It is therefore possible that a significant 
number of Population III stars may form in such conditions.

A final possibility is that the required ionization can be produced by a flux of X-rays or high energy cosmic rays.
Although X-ray ionization was initially favoured as a means of raising the ionization level of the gas, and hence
promoting H$_{2}$ formation \citep{har00}, more recent work has shown that if one considers realistic models 
for the X-ray background that also account for the simultaneous growth of the soft UV background, then one finds
that UV photodissociation of the H$_{2}$ is a more important effect, and hence that the growth of the radiation \index{photodissociation}
backgrounds almost always leads to an overall reduction in the amount of H$_{2}$ produced \citep{gb03,mba03}.
Cosmic ray ionization may therefore prove to be the more important effect \citep{sb07,jce07}, although we still
know very little about the likely size of the cosmic ray ionization rate in high redshift minihalos. \index{molecular hydrogen!cooling|)}
 \index{HD!cooling|)}

\subsubsection{Three-body H$_{2}$ formation}
\index{molecular hydrogen!chemistry|(}
\index{molecular hydrogen!three-body reaction|(}
Once the collapsing gas reaches a density of around $10^{8}$--10$^{9} \: {\rm cm^{-3}}$, 
its chemical makeup starts to change significantly. The reason for this is that at these densities, the formation of
H$_{2}$ via the three-body reactions \citep{pss83}
\begin{eqnarray}
{\rm H} + {\rm H} + {\rm H} & \rightarrow & {\rm H_{2}} + {\rm H}, \label{3b1} \\
{\rm H} + {\rm H} + {\rm H_{2}} & \rightarrow & {\rm H_{2}} + {\rm H_{2}}, \\
{\rm H} + {\rm H} + {\rm He} & \rightarrow & {\rm H_{2}} + {\rm He},
\end{eqnarray}
starts to become significant. These reactions quickly convert most of the hydrogen in the gas into H$_{2}$. At the same time,
however, they generate a substantial amount of thermal energy: every time an H$_{2}$ molecule forms via one of these
three-body reactions, its binding energy of 4.48~eV is converted into heat. A simple estimate of the relative sizes of
the compressional heating rate and the three-body H$_{2}$ formation heating rate helps to demonstrate the importance
of the latter during this stage of the collapse. Let us consider gas at a density $n = 10^{8} \: {\rm cm^{-3}}$ that has a 
temperature $T = 1000 \: {\rm K}$, collapsing at a rate such that ${\rm d}n/{\rm dt} = n / t_{\rm ff}$, where $t_{\rm ff}$ is
the gravitational free-fall time. In these conditions, the compressional heating rate is given by
\begin{eqnarray}
\Lambda_{\rm pdv} & \simeq & 1.25 \times 10^{-31} n^{1/2} T, \\
 & = & 1.25 \times 10^{-24} \: {\rm erg} \: {\rm s^{-1}} \: {\rm cm^{-3}},
\end{eqnarray}
while the three-body H$_{2}$ formation heating rate has the value
\begin{eqnarray}
\Lambda_{\rm 3b} & \simeq & 3.9 \times 10^{-40} T^{-1} n^{3} x_{\rm H}^{3}, \\
 & = & 3.9 \times 10^{-19} x_{\rm H}^{3} \: \: {\rm erg} \: {\rm s^{-1}} \: {\rm cm^{-3}},
\end{eqnarray}
where we have adopted the rate coefficient for reaction~\ref{3b1} given in \citet{pss83}. Comparing the two heating rates,
we see that three-body H$_{2}$ formation heating dominates unless $x_{\rm H}$ is very small (i.e.\ unless the gas is almost
fully molecular). Therefore, even though the abundance of H$_{2}$, the dominant coolant during this phase of the collapse, 
increases by more than two orders of magnitude, the gas typically does not cool significantly, owing to the influence of this 
three-body H$_{2}$ formation heating. Indeed, the temperature often actually increases.

One major uncertainty that remains in current treatments of this phase of the collapse of the gas is exactly how quickly
the gas becomes molecular. Although reaction~\ref{3b1} is the dominant source of H$_{2}$ at these densities, the
rate coefficient for this reaction is poorly known, with published values differing by almost two orders of magnitude
at 1000~K, and by an even larger factor at lower temperatures \citep{g08,turk11}. The effects of this uncertainty have 
recently been studied by \citet{turk11}. They show that it has little effect on the density profile of the gas, and only a
limited effect on the temperature profile. However, it has much more significant effects on the morphology of the gas
and on its velocity structure. Simulations in which a high value was used for the three-body rate coefficient show
find that gas occurs more rapidly, and that the molecular gas develops a much more flattened, filamentary structure.
Significant differences are also apparent in the infall velocities and the degree of rotational support. \citet{turk11}
halt their simulations at the point at which a protostar first forms, and so do not directly address the issue of whether
these differences continue to have an influence during the accretion phase, and whether the affect the ability of the
gas to fragment (see Section~\ref{fragment} below). A follow-up study that focussed on these issues would be 
informative.
\index{molecular hydrogen!chemistry|)}
\index{molecular hydrogen!three-body reaction|)}

\subsubsection{Optically-thick line cooling}
\index{molecular hydrogen!cooling|(}
The next important event occurs at a density of around $10^{10} \: {\rm cm^{-3}}$, 
when the main rotational and vibrational lines of H$_{2}$ start to become optically thick \citep{ra04}. The effect of this 
is to reduce the efficiency of H$_{2}$ cooling, leading to a continued rise in the gas temperature. In one-dimensional
simulations \citep[e.g.][]{on98,onus98,ripa02}, it is possible to treat optically thick H$_{2}$ cooling accurately by solving the
full radiative transfer problem.  These models show that although the optical depth of the gas becomes large at frequencies 
corresponding to the centers of the main H$_{2}$ emission lines, the low continuum opacity of the gas allows photons
to continue to escape through the wings of the lines, with the result that the H$_{2}$ cooling rate is suppressed far
less rapidly as the collapse proceeds than one might at first expect \citep[see][for a detailed discussion of this point]{onus98}.

In three-dimensional simulations, solution of the full radiative transfer problem is not currently possible, due to the
high computational expense, which has motivated a search for simpler approximations. There are two such approximations
in current use in simulations of Population III star formation. The first of these was introduced by \citet{ra04}. They
proposed that the ratio of the optically thick and optically thin H$_{2}$ cooling rates,
\begin{equation}
f_{\tau} \equiv \frac{\Lambda_{\rm H_{2}, thick}}{\Lambda_{\rm H_{2}, thin}},
\end{equation}
could be represented as a simple function of density:
\begin{equation}
f_{\tau} = {\rm min} \left[1, \left(\frac{n}{n_{0}}\right)^{-0.45} \right],
\end{equation}
where $n_{0} = 8 \times 10^{9} \: {\rm cm^{-3}}$. They showed that this simple expression was a good approximation to 
the results of the full radiative transfer model used by \citet{ripa02}, and suggested that this approximation would be useful
for extending the results of three-dimensional simulations into the optically thick regime. However, they also noted that it
may only be accurate while the collapse remains approximately spherical, as the one-dimensional model on which it is
based assumes spherical infall.

An alternative approach was introduced by \citet{yoha06}. They compute escape probabilities for each rotational and
vibrational line of H$_{2}$ using the standard Sobolev \index{Sobolev approximation|(}
approximation \citep{sobolev}. In this approximation, the optical
depth at line centre of a transition from an upper level $u$ to a lower level $l$ is written as
\begin{equation}
\tau_{ul} = \alpha_{ul} L_{\rm s},
\end{equation}
where $\alpha_{ul}$ is the line absorption coefficient and $L_{\rm s}$ is the Sobolev length. The absorption coefficient
$\alpha_{ul}$ can be written as
\begin{equation}
\alpha_{ul} = \frac{\Delta E_{ul}}{4\pi} n_{l} B_{lu} \left[1 - \exp \left(\frac{-\Delta E_{ul}}{kT} \right) \right] \phi(\nu_{ul}),
\end{equation}
where $E_{ul}$ is the energy difference between the two levels, $n_{l}$ is the number density of H$_{2}$ molecules in the 
lower levels, $B_{lu}$ is the usual Einstein B coefficient, and $\phi(\nu_{ul})$ is the line profile at the centre of the line.
The Sobolev length is given by 
\begin{equation}
L_{\rm s} = \frac{v_{\rm th}}{|{\rm d}v_{\rm r} / {\rm d}r |},
\end{equation}
where $v_{\rm th}$ is the thermal velocity of the H$_{2}$ and $|{\rm d}v_{\rm r} / {\rm d}r |$ is the size of the velocity gradient
along a given line of sight from the fluid element of interest. Given $\tau_{ul}$, the escape probability for photons emitted in
that direction then follows as
\begin{equation}
\beta_{ul} = \frac{1 - \exp(-\tau_{ul})}{\tau_{ul}}.
\end{equation}
To account for the fact that the velocity gradient may differ along different lines of sight from any particular fluid element,
\citet{yoha06} utilize a mean escape probability given by
\begin{equation}
\beta = \frac{\beta_{x} + \beta_{y} + \beta_z}{3},
\end{equation}
where $\beta_{x}$, $\beta_{y}$ and $\beta_{z}$ are the escape probabilities along lines of sight in the $x$, $y$ and $z$
directions, respectively. Finally, once the escape probabilities for each transition have been calculated, the optically thick
H$_{2}$ cooling rate can be computed from
\begin{equation}
\Lambda_{\rm H_{2}, thick} = \sum_{u, l} E_{ul} \beta_{ul} A_{ul} n_{u},
\end{equation}
where $A_{ul}$ is the Einstein A coefficient for the transition from $u$ to $l$ and $n_{u}$ is the population of the upper
level $u$. 

Strictly speaking, the Sobolev approximation is valid only for flows in which the Sobolev length $L_{\rm s}$ is
much smaller than the characteristic length scales associated with changes in the density, temperature or
chemical makeup of the gas, a requirement which is easy to satisfy when the velocity gradient is very large,
but which is harder to justify in the case of Population III star formation, since the collapse speed is typically
comparable to the sound-speed.  Nevertheless, \citet{yoha06} show that the optically thick H$_{2}$ cooling
rates predicted by the Sobolev approximation are in very good agreement with those computed in the 
one-dimensional study of \citet{on98} by solution of the full radiative transfer problem. 
\index{Sobolev approximation|)}

Little work has been done on comparing these two approaches to treating optically-thick H$_{2}$ cooling.
This issue was addressed briefly in \citet{turk11}, who showed that the two approximations yielded similar
values for $f_{\tau}$ for densities $n < 10^{15} \: {\rm cm^{-3}}$ during the initial collapse of the gas,
with differences of at most a factor of two. However, as yet no study has examined whether this good 
agreement persists past the point at which the first protostar forms.

\subsubsection{Collision-induced emission}
\index{molecular hydrogen!collision-induced emission|(}
A further significant point in the collapse of the gas is reached once the number density increases to
$n \sim 10^{14} \: {\rm cm^{-3}}$. At this density, a process known as collision-induced emission 
becomes important. Although an isolated H$_{2}$ molecule has no dipole moment, and can only
emit or absorb radiation through quadrupole transitions, when two H$_{2}$ molecules collide\footnote{
A similar process can also occur during collisions of atomic hydrogen or atomic helium with H$_{2}$,
but it is the H$_{2}$-H$_{2}$ case that is the most relevant here} they briefly act as a kind of
``supermolecule'' with a non-zero dipole moment for the duration of the collision. This supermolecule
can therefore absorb or emit radiation through dipole transitions, which have much higher transition
probabilities than the quadrupole transitions available to isolated H$_{2}$. If radiation is absorbed,
this process is termed collision-induced absorption; if it is emitted, then we refer to the process as
collision-induced emission (CIE). A detailed discussion of the phenomenon can be found in
\citet{fromm93}.

Collision-induced emission can in principle occur in gas of any density, but the probability of a photon
being emitted in any given collision is very small, owing to the short lifetime of the collision state 
($\Delta t < 10^{-12} \: {\rm s}$ at the temperatures relevant for Pop.\ III star formation; see \citealt{ra04}).
For this reason, CIE becomes an important process only at very high gas densities. Another consequence
of the short lifetime of the collision state is that the individual lines associated with the dipole transitions
become so broadened that they actually merge into a continuum. This is important, as it means that the
high opacity of the gas in the rovibrational lines of H$_{2}$ does not significantly reduce the amount
of energy that can be radiated away by CIE. Therefore, once the gas reaches a sufficiently high density,
CIE becomes the dominant form of cooling, as pointed out by several authors \citep{on98,ripa02,ra04}.

The most detailed study of the effects of CIE cooling on the collapse of primordial gas was carried out
by \citet{ra04}. They showed that CIE cooling could actually become strong enough to trigger a thermal
instability, However, the growth rate of this instability is longer than the gravitational free-fall time, meaning
that it is unlikely that this process can drive fragmentation during the initial collapse of the gas. 
\index{molecular hydrogen!collision-induced emission|)}

\subsubsection{Cooling due to H$_{2}$ dissociation}
\index{molecular hydrogen!chemistry|(}
The phase of the collapse dominated by CIE cooling lasts for only a relatively short period of time. The
gas becomes optically thick in the continuum once it reaches a density  $ n \sim 10^{16} \: {\rm cm^{-3}}$
\citep{on98,ra04}, which strongly suppresses any further radiative cooling. Once this occurs, the gas
temperature rises until it reaches a point at which the H$_{2}$ begins to dissociate. At these densities,
this occurs at a temperature $T \sim 3000$~K. Once
this point  is reached, the temperature rise slows, as most of the
energy released during the collapse goes into dissociating the H$_{2}$ rather than raising the temperature.
As it takes 4.48~eV of energy to destroy each H$_{2}$ molecule, this H$_{2}$ dissociation phase continues 
for a while. However, it comes to an end once almost all of the H$_{2}$ has been destroyed, 
at which point the temperature of the gas begins to climb steeply. The thermal 
pressure in the interior of the collapsing core rises rapidly and eventually becomes strong enough to halt the 
collapse. At the point at which this occurs, the size of the dense core is around 0.1~AU, its mass is around
$0.01 \: {\rm M_{\odot}}$ and its mean density is of order $10^{20} \: {\rm cm^{-3}}$ \citep{yoh08}. It is bounded 
by a strong accretion shock. This pressure-supported, shock-bounded core is a Population III protostar, and
its later evolution is discussed in Section~\ref{later} below.
\index{molecular hydrogen!chemistry|)}
\index{molecular hydrogen!cooling|)}

\subsection{Dark matter annihilation}
\label{dma}
One complication not accounted for in the models of Pop.\ III star formation described above is the role
that may be played by dark matter annihilation. The nature of dark matter is not yet understood, but one
plausible candidate is a weakly interacting massive particle (WIMP) -- specifically,  the lightest  \index{WIMP}
supersymmetric  particle predicted in models based on supersymmetry. The simplest supersymmetry models 
predict that this WIMP  has an annihilation cross-section $\langle\sigma v \rangle
\sim 3 \times 10^{-26} \: {\rm cm^{2}}$, a mass within the range of 50~GeV to 2~TeV, and a cosmological
density consistent with the inferred density of dark matter \citep{spol08}. The rate per unit volume at which 
energy is produced by dark matter annihilation can be written as $Q_{\rm ann} = \langle\sigma v \rangle \rho_{X}^{2} 
/ m_{X}$, where $\rho_{X}$ is the mass density of dark matter and $m_{X}$ is the mass of a single dark
matter particle. For a plausible particle mass of 100~GeV, and a dark matter density equal to the cosmological
background density of dark matter, this expression yields a tiny heating rate, $Q_{\rm ann} \sim 6 \times 10^{-62}
(1+z)^{6} \Omega_{\rm m}^{2} h^{4} \: {\rm erg} \: {\rm cm^{-3}} \: {\rm s^{-1}}$, even before one accounts for the
fact that much of the annihilation energy is released in the form of energetic neutrinos or gamma-rays that couple
only very weakly with the intergalactic gas. WIMP annihilation therefore plays no significant role in the evolution \index{WIMP}
of the intergalactic medium while the WIMPs remain uniformly distributed \citep{mn08}. However, the $\rho_{X}^{2}$ 
density dependence of the heating rate means that it can potentially become significant in regions where the dark 
matter density is very high.

\citet{spol08} proposed that one situation in which the heating from dark matter annihilation could
become important would occur during the formation of the very first Population III protostars. They assumed
that any given star-forming minihalo would form only a single Pop.\ III protostar, and that this protostar would
form at the center of the minihalo. As the gas collapsed at the center of the minihalo, its increasing gravitational
influence would bring about a local enhancement of the dark matter density, via a process known as adiabatic
contraction. \index{adiabatic contraction}
The basic idea underlying this is very simple. For a collisionless particle on a periodic orbit, the 
quantity $\oint p dq$, where $p$ is the conjugate momentum of coordinate $q$, is an adiabatic invariant, i.e.\ a 
quantity that does not vary when the gravitational potential varies, provided that the rate of change of the potential
is sufficiently slow. If $p$ represents the angular momentum of a particle on a circular orbit of radius $r$ within some
spherically symmetric mass distribution, then one can show that the quantity $r M(r)$ is constant for that particle,
where $M(r)$ is the mass enclosed within $r$, so long as this enclosed mass changes on a timescale that is
long compared to the orbital period. \citet{spol08} show that if one starts with a simple NFW profile for the dark matter
\citep{nfw97} and account for the effects of adiabatic contraction using a simple approach pioneered by
\citet{blum86}, then one finds that for any WIMP mass less than 10~TeV, the effects of dark matter  \index{WIMP}
annihilation heating become significant during the collapse of the gas. \citet{spol08} identify the point at 
which this occurs by comparing the heating rate due to dark matter annihilation with the H$_{2}$ cooling \index{molecular hydrogen!cooling}
rate. To determine a value for the latter, they make use of the simulation results of  \citet{yoha06} and \citet{gao07}
and measure how the H$_{2}$ cooling rate of the gas in the central collapsing core evolves as the collapse
proceeds. They show that
for a 100~GeV WIMP, heating dominates at gas densities $n > 10^{13} \: {\rm cm^{-3}}$. Finally, they argue
that once dark matter annihilation heating dominates over H$_{2}$ cooling, the gravitational collapse of the
gas will come to a halt, and hence the gas will never reach protostellar densities. Instead, it will remain
quasi-statically supported at a density of roughly $10^{13} \: {\rm cm^{-3}}$ (for  a 100~GeV WIMP), with  \index{WIMP}
a corresponding size scale of 17~AU, for as long as the dark matter annihilation rate remains large
compared to the H$_{2}$ cooling rate. As the time required to consume all of the dark matter within a radius
of 17~AU may be hundreds of millions of years, the resulting quasi-static gas distribution -- dubbed a ``dark star'' \index{dark star}
by \citet{spol08} --  could potentially survive for a very long time.

One criticism of the original \citet{spol08} model is its reliance on the  \citet{blum86} prescription for describing 
the effects of the adiabatic contraction of the dark matter.  \index{adiabatic contraction}
This prescription assumes that all of the dark matter
particles move on circular orbits, which is unlikely to be the case in a realistic dark matter minihalo, and concerns
have been expressed that it may yield values for the dark matter density after adiabatic contraction that are
significantly higher than the true values \citep[see e.g.][]{gn04}. For this reason, \citet{freese09} re-examined this 
issue using an alternative method for estimating the effects of adiabatic contraction, based on \citet{young80}. 
This alternative prescription does account for particles moving on radial orbits, and \citet{freese09} show that it 
predicts dark matter densities that are indeed systematically smaller than those predicted by the \citet{blum86} 
prescription, but only by a factor of two. \citet{freese09} therefore conclude that  although using the \citet{young80} 
prescription for adiabatic contraction in place of the simpler \citet{blum86} prescription will lead to some minor
quantitative changes in the predicted outcome, the main qualitative results of the \citet{spol08} study are 
insensitive to this change, and one would still expect a ``dark star'' to form. \index{dark star}
 
\index{molecular hydrogen!cooling|(}
Another potential problem with the dark star hypothesis is the fact that it is not at all clear that the collapse of
the gas will stop once the dark matter heating rate exceeds the H$_{2}$ cooling rate. For one thing, the values
for the H$_{2}$ cooling rate used by \citet{spol08} do not account for the effects of the dark matter annihilation
heating. If this leads to an increase in temperature, then this will also increase the H$_{2}$ cooling rate, 
allowing more of the energy produced by  dark matter annihilation to be radiated away. It is therefore unlikely
that the point in the collapse at which the dark matter annihilation heating rate exceeds the Spolyar et al.\
estimate for the H$_{2}$ cooling rate is marked by any sharp jump in the temperature. Instead, we would 
expect to find a more gradual temperature increase, at least up until the point at which collisional dissociation 
of the H$_{2}$ starts to occur. 

Once H$_{2}$ begins to dissociate, this provides another outlet for the energy generated by dark matter
annihilation. Spolyar et al.\ estimate that for a 100~GeV WIMP, the power generated by dark matter  \index{WIMP}
annihilation within the central core is $L_{\rm dm} \sim 140 \: {\rm L_{\odot}}$, and the core mass is
roughly $0.6 \: {\rm M_{\odot}}$. The total energy stored within the core in the form of the binding
energy of the H$_{2}$ molecules is roughly
\begin{equation}
E_{\rm H_{2}, bind}  =  4.48 {\rm eV} \times 0.76 \times \frac{0.6 \: {\rm M_{\odot}}}{m_{\rm H_{2}}} 
  \simeq  2.6 \times 10^{45} \: {\rm erg},
\end{equation}
and the time required for dark matter annihilation to produce this much energy is
\begin{equation}
t_{\rm dis} = \frac{E_{\rm H_{2}, bind}}{L_{\rm dm}} \simeq 200 \: {\rm yr}.
\end{equation}
For comparison, the free-fall time at this point in the collapse is roughly 15~years. H$_{2}$ dissociation
will therefore allow the collapse of the gas to continue until either the dark matter heating rate becomes
large enough to destroy the H$_{2}$ in the core in much less than a dynamical time, or the compressional
heating produced during the collapse becomes capable of doing the same job. In either case, it is likely
that much higher core densities can be reached than was assumed in the Spolyar et~al.\ study.

A first attempt to hydrodynamically model the formation of a ``dark star'' while correctly accounting for \index{dark star}
these thermodynamical effects was made by \citet{ripa10}. They used the 1D, spherically symmetric
hydrodynamical code described in \citet{ripa02} to model the collapse of the gas up to densities of 
order $10^{15} \: {\rm cm^{-3}}$ for a range of different WIMP masses between 1~GeV and 1~TeV.  \index{WIMP}
Adiabatic contraction of the
dark matter was modelled using the algorithm described in \citet{gn04}, and the effects of the dark
matter annihilation heating and ionization were self-consistently accounted for in the chemical
and thermal model. \citet{ripa10} show that even in the most extreme case that they study, the heating
produced by the dark matter appears unable to halt the collapse for an extended period. After the
dark matter heating rate exceeds the H$_{2}$ cooling rate, dissociation of H$_{2}$ in the core accounts
for most of the ``excess'' energy not radiated away by the gas, allowing the collapse to continue. Once
the H$_{2}$ in the core is exhausted, the temperature rises steeply, very briefly halting the collapse.
However, the temperature quickly becomes large enough to allow other cooling mechanisms (e.g.\
H$^{-}$ bound-free transitions or Lyman-$\alpha$ emission from atomic hydrogen) to operate, allowing
the collapse to restart. \citet{ripa10} do not find any evidence for the formation of a hydrostatically
supported ``dark star'' up to the highest densities that they study. Confirmation of this result in a 3D
treatment of the collapse would be very useful. \index{dark star}
\index{molecular hydrogen!cooling|)}
 
\subsection{The role of magnetic fields}
\label{magn}
\subsubsection{Initial strength}
\index{magnetic fields|(}
\index{magnetic fields!initial strength|(}
The majority of the work that has been done on modelling the formation of the first stars assumes that magnetic fields
play no role in the process, either because no magnetic field exists at that epoch, or because the strength of any field 
that does exist is too small to be significant. A number of mechanisms have been suggested that may generate 
magnetic seed fields during the inflationary epoch, the electroweak phase transition or the QCD phase transition
\citep[see][for a recent comprehensive review]{kandus10}. Observational constraints \citep[e.g.][]{bfs97,sbk08}
limit the strength of the magnetic field at the epoch of first star formation to no more than about 1~nG (in comoving units), 
but it is quite possible that any primordial seed field resulting from one of these processes will actually have a much smaller 
strength.

An alternative source for magnetic fields within the first generation of star-forming minihalos is the so-called
Biermann battery effect \citep{biermann50}. In a partially ionized gas in which the gradient of electron 
density \index{magnetic fields!Biermann battery|(}
does not perfectly align with the gradient of electron pressure, as can happen if there is a temperature gradient
that does not align with the pressure gradient, the magnetic induction equation takes the form 
\begin{equation}
\frac{\partial {\mathbf B}}{\partial t} = \nabla \times ({\mathbf v} \times {\mathbf B}) + \frac{c \nabla p_{\rm e} \times \nabla n_{\rm e}}
{n_{\rm e}^{2} e}, 
\end{equation}
where ${\mathbf B}$ is the magnetic field, ${\mathbf v}$ is the velocity, $n_{\rm e}$ is the electron density, $p_{\rm e}$ is the electron pressure, 
and $e$ is the charge on an electron. In the limit that $B \rightarrow 0$, the first term on the right-hand side 
of this equation also becomes zero, but the battery term does not. It can therefore act as the source of a magnetic field
in a gas that is initially unmagnetized. An early investigation into the effectiveness of the Biermann battery during
galaxy formation was made by \citet{kuls97}, who considered the formation of massive galaxies and showed that
the Biermann battery could generate a field of strength $B \sim 10^{-21}$~G during their assembly. More recently, 
\citet{xu08} have simulated the action of the Biermann battery during the formation of one of the first star-forming
minihalos, finding that it is able to generate initial field strengths of the order of $10^{-18} \: {\rm G}$ during this
process. \index{magnetic fields!Biermann battery|)}
\index{magnetic fields!initial strength|)}

\subsubsection{Amplification}
\index{magnetic fields!amplification|(}
The seed fields generated by the Biermann battery, or by other processes acting in the very early Universe can
be significantly amplified by flux-freezing during the gravitational collapse of the gas. If the diffusive timescale
associated with ambipolar diffusion or Ohmic diffusion is long compared to the gravitational collapse timescale,
then the magnetic field will be ``frozen'' to the gas, and will be carried along with it when the gas collapses. 

In the optimal case of spherical collapse, perfect flux freezing implies that the field strength evolves with density
as $B \propto \rho^{2/3}$, and hence the magnetic pressure $p_{\rm mag} = B^{2} / 8 \pi$ evolves as $p_{\rm mag}
\propto \rho^{4/3}$. In comparison, the thermal pressure $p_{\rm therm}$ evolves as $p_{\rm therm} \propto
\rho T$, and so if the temperature does not vary much during the collapse, the plasma $\beta$ parameter,
$\beta \equiv p_{\rm therm} / p_{\rm mag}$, evolves as $\beta \propto \rho^{-1/3}$. Therefore, if the gas is
initially dominated by thermal pressure rather than magnetic pressure, it will remain so during much of the
collapse, as a large change in the density is necessary to significantly alter $\beta$.
In the case examined by \citet{xu08}, the very small initial magnetic field strength means that
$\beta$ is initially very large, and remains so throughout the collapse, implying that the magnetic field 
never becomes dynamically significant. Moreover, even if we take an initial comoving field strength of 1~nG,
comparable to the observational upper limit, at the mean halo density, $\beta \sim 10^{4}$ (assuming a
halo formation redshift $z=20$ and a virial temperature of 1000~K), and does not become of order unity
until very late in the collapse. Furthermore, if the collapse of the gas is not spherical, whether because of
the effects of gravitational forces, angular momentum, or the influence of the magnetic field itself, the
amplification due to flux freezing and collapse will be less than in the spherical case \citep[see e.g.][who 
find a somewhat shallower relationship in some of their models]{mach06}.

Therefore, for magnetic fields  to play an important role in Pop.\ III star formation, they must either start with
a field strength very close to the observational upper limit, or we must invoke an amplification process that
is much more effective than the amplification that occurs due to flux freezing and gravitational collapse.
One obvious possibility is amplification via some kind of dynamo process, which could bring about
exponential amplification of an initially small seed field. Of particular interest is the small-scale turbulent
dynamo \citep{kn67,kaz68,ka92}. This produces a magnetic field that has no mean flux on the 
largest scales but that can have substantial mean flux within smaller-scale subregions. The growth rate of the 
magnetic field due to the turbulent dynamo is closely related to the rate of turnover of the smallest eddies. If 
the magnetic field is sufficiently small that it does not significantly affect the velocity field of the gas (the 
kinematic approximation), and if we assume that we are dealing with Kolmogorov turbulence, then
\citet{kz08} show that the magnetic energy density grows exponentially, and that after a single gravitational
free-fall time it is amplified by a factor $\exp({\rm Re}^{1/2})$, where ${\rm Re}$ is the Reynolds number
of the flow. If we assume that the driving scale of the turbulence is comparable to the size of the minihalo,
and that the turbulent velocity is of the same order as the sound speed \citep[see e.g.][]{abn02}, then 
${\rm Re} \sim 10^{4}$--$10^{5}$, implying that the magnetic field is amplified by an enormous factor
during the collapse. In practice, the field will not be amplified by as much as this analysis suggests, as
the kinematic approximation will break down once the magnetic energy density becomes comparable
to the kinetic energy density on the scale of the smallest eddies. Nevertheless, this simple treatment
implies that the turbulent dynamo can amplify the magnetic field to a strength at which it becomes
dynamically important.

Although the importance of dynamo processes during the formation of the first galaxies has been
understood for a number of years \citep[see e.g.][]{ps89,beck94,kuls97}, they have attracted surprisingly
little attention in studies of primordial star formation. Over the past couple of years, however, this
has begun to change, with several recent studies focussing on the growth of magnetic fields during
the formation of the first stars. The first of these was \citet{sch10}, who studied the
effectiveness of the turbulent dynamo during gravitational collapse using a simple one-zone
Lagrangian model for the collapsing gas. Their model assumes that turbulence is generated
by gravitational collapse on a scale of the order of the Jeans length, and that on smaller scales, the  \index{Jeans length}
turbulent velocity scales with the length-scale $l$ as $v_{\rm turb} \propto l^{\beta}$. \citet{sch10}
study both Kolmogorov turbulence, with $\beta = 1/3 $ and Burgers turbulence, with $\beta = 1/2$,
and show that in both cases, amplification of a weak initial seed field occurs rapidly, and that the
field reaches saturation on all but the largest scales at an early point during the collapse. 
Because \citet{sch10} did not solve directly for the fluid velocities, they were unable to model
the approach to saturation directly. Instead, they simply followed \citet{subra98} and assumed that 
the strength of the saturated field satisfies
\begin{equation}
B_{\rm sat} = \sqrt{\frac{4 \pi \rho v_{\rm turb}^{2}}{{\rm Rm_{cr}}}},
\end{equation}
where ${\rm Rm_{cr}} \sim 60$ is a critical value of the magnetic Reynolds number, 
${\rm Rm} = v_{\rm turb} l / \eta$ (where $\eta$ is the resistivity), that must be exceeded in
order for exponential growth of the field to occur \citep{subra98}.

The main weakness of the \citet{sch10} study lies in the assumptions that it was forced to make
about the nature of the turbulent velocity field. Therefore, in a follow-up study, \citet{sur10} used 
high-resolution adaptive mesh refinement simulations to directly follow the coupled evolution of 
the velocity field and the magnetic field within a 3D collapse model. For their initial conditions,
\citet{sur10} took a super-critical Bonnor-Ebert sphere \citep{bonnor56,ebert55} with a core
density $n_{\rm c} = 10^{4} \: {\rm cm^{-3}}$ and a temperature $T = 300 \: {\rm K}$. They
included initial solid-body rotation, with a rotational energy that was 4\% of the total gravitational
energy, and a turbulent velocity component with an RMS velocity equal to the sound speed and 
with an energy spectrum $E(k) \propto k^{-2}$. A weak magnetic field was also included, with
an RMS field strength $B_{\rm rms} = 1 \: {\rm nG}$, and with the same energy spectrum as the
turbulence. For reasons of computational efficiency, \citet{sur10} did not follow the thermal and
chemical evolution of the gas directly. Instead, they adopted a simple barotropic equation of 
state, $P \propto \rho^{1.1}$, inspired by the results of previous hydrodynamical models
\citep[e.g.][]{abn02}. In view of the sensitivity of the turbulent dynamo to the Reynolds number,
and the fact that numerical dissipation on the grid scale limits the size of Re in any 3D 
numerical simulation to be substantially less than the true physical value, there is good reason
to expect that the dynamo amplification rate will be sensitive to the numerical resolution of the
simulation. \citet{sur10} therefore focussed on the effects of resolution, performing five different
simulations with the same initial conditions, but with different grid refinement criteria. Starting
with a model in which the refinement criterion ensures that the Jeans length is always resolved  \index{Jeans length}
by 8 grid zones, they looked at the effects of increasing this number to 16, 32, 64 and 128 grid zones.

\citet{sur10} showed that in the 8 and 16 cell runs, the magnetic field strength increases with density
at a slower rate than the $B \propto \rho^{2/3}$ that we would expect simply from flux freezing and
roughly spherical collapse, indicating that in these runs, the turbulent dynamo does not operate.
Starting with the 32 cell run, however, they found evidence for an increase in $B$ with density that
is larger than can be explained simply by compression, which they ascribe to the effects of the
turbulent dynamo. They showed that as the number of grid zones used to resolve the Jeans length  \index{Jeans length}
is increased, the rate at which the field grows also increases, and there is no sign of convergence
at even their highest resolution. This resolution dependence explains why the earlier study of
\citet{xu08} found no evidence for dynamo amplification, as their study used only 16 grid zones
per Jeans length. 

More recently, \citet{fed11} have re-examined this issue of resolution dependence, and have 
shown that when the number of grid zones per Jeans length is small, the amount of turbulent  \index{Jeans length}
energy on small scales is also significantly underestimated. The reason for this is the same as
the reason for the non-operation of the turbulent dynamo: the effective Reynolds number is   \index{Reynolds number|(}
too small. \citet{fed11} show that in gravitationally collapsing regions that undergo adaptive
mesh refinement, the effective Reynolds number scales with the number of grid zones per
Jeans length as ${\rm Re_{eff}} = (N / 2)^{4/3}$. Furthermore, an effective Reynolds number
${\rm Re_{eff}} \sim 40$ is required in order for the turbulent dynamo to operate, implying
that one needs $N \sim 30$ or more zones per Jeans length in order to begin resolving it,
in agreement with the findings of \citet{sur10}. It should also be noted that the operation
of the turbulent dynamo in simulations of turbulence without self-gravity requires a similar
minimum value for the Reynolds number \citep{haugen04}.  \index{Reynolds number|)}

Together, these studies support the view that amplification of a weak initial magnetic field by 
the turbulent dynamo may indeed have occurred within the first star-forming minihalos. 
However, a number of important issues remain to be addressed. First, the three-dimensional
studies carried out so far all adopt a simple barotropic equation of state, rather than solving
self-consistently for the thermal evolution of the gas. This is a useful simplifying assumption,
but may lead to incorrect dynamical behaviour, as one misses any effects due to thermal
instabilities, or the thermal inertia of the gas (i.e.\ the fact that the cooling time is typically
comparable in size to the dynamical time). Work is currently in progress to re-run some of
these models with a more realistic treatment of the thermodynamics and chemistry in order
to explore the effect that this has on the degree of amplification (T.\ Peters, private communication).
Second, it will clearly be important to perform similar studies using more realistic initial conditions
for the gas. Of particular concern is whether the turbulence that is generated during the gravitational
collapse of gas within a primordial minihalo is similar in nature to that studied in these more
idealized calculations, and if not, what influence this has on the amplification of the field.
Finally, and most importantly, there is the issue of the level at which the field saturates.
Exponential amplification of the field by the small-scale dynamo will occur only while the
kinematic approximation holds, i.e.\ while the energy stored in the magnetic field is much
smaller than the energy stored in the small-scale turbulent motions. Once the field becomes
large, the Lorentz force that it exerts on the gas will act to resist further folding and amplification
of the field. In addition, the dissipation of magnetic energy by Ohmic diffusion and ambipolar
diffusion will grow increasingly important. However, it remains unclear which of these effects
will be the most important for limiting the growth of the magnetic field in dense primordial gas.
\index{magnetic fields!amplification|)}

\subsubsection{Consequences}
\index{magnetic fields!effects|(}
If a strong magnetic field can be generated by dynamo amplification, then it will affect both
the thermal and the dynamical evolution of the gas. The possible dynamical effects of a
strong magnetic field have been investigated by \citet{mach06,mach08}. In a preliminary 
study, \citet{mach06} used nested-grid simulations to investigate the influence of a magnetic
field on the collapse of a small, slowly-rotating primordial gas cloud. For their initial conditions,
they used a supercritical Bonnor-Ebert sphere  with mass $5.1 \times 10^{4}
\: {\rm M_{\odot}}$, radius $6.6 \: {\rm pc}$, central density $n_{\rm c} = 10^{3} \: {\rm cm^{-3}}$
and an initial temperature of 250~K. They assumed that this cloud was in solid body rotation
with angular velocity $\Omega_{0}$ and that it was threaded by a uniform magnetic field
oriented parallel to the rotation axis, with an initial field strength $B_{0}$. They performed
simulations with several different values of $\Omega_{0}$ and $B_{0}$, with the former
ranging from $10^{-17} \: {\rm s^{-1}}$ to $3.3 \times 10^{-16} \: {\rm s^{-1}}$, and the latter
from $10^{-9} \: {\rm G}$ to $10^{-6} \: {\rm G}$. To treat the thermal evolution of the gas, they
used a barotropic equation of state, based on the one-zone results of \citet{om05}.

\citet{mach06} used this numerical setup to follow the collapse of the gas down to scales of
the order of the protostellar radius. They showed that the magnetic field was significantly
amplified by compression and flux freezing during the collapse, reaching strengths of order
$6 \times 10^{5}$--$6 \times 10^{6} \: {\rm G}$ on the scale of the protostar. A very compact
disk with a radius of few R$_{\odot}$ formed around the protostar, and in models with
initial field strength $B_{0} > 10^{-9} \: {\rm G}$, the magnetic field became strong enough
to drive a hydromagnetic disk wind that ejected roughly 10\% of the infalling gas. 
Numerical limitations (discussed in Section~\ref{later} below) prevented \citet{mach06} from 
following the evolution of the system for longer than a few days after the formation of the
protostar, and so it remains unclear whether an outflow would eventually be generated
in the $10^{-9} \: {\rm G}$ case, and whether the outflows continue to be driven as
the protostar and disk both grow to much larger masses.

In a follow-up study, \citet{mach08} used a similar numerical setup, but examined a much
broader range of values for $\beta_{0} (\equiv E_{\rm rot} / |E_{\rm grav}|)$, the ratio of
the initial rotational energy to the initial gravitational energy, and $\gamma_{0} (\equiv
E_{\rm mag} / |E_{\rm grav}|)$, the ratio of the initial magnetic energy to the initial gravitational 
energy. They found that the outcomes of the simulations could be classified into two main
groupings. Clouds with $\beta_{0} > \gamma_{0}$, i.e.\ ones which were rotationally
dominated, formed a prominent disk during the collapse that then fragmented into a binary
or higher order multiple system. In these simulations, no jets were seen (with the exception
of  a couple of model in which $\beta_{0} \sim \gamma_{0}$). On the other hand, when
$\beta_{0} < \gamma_{0}$, i.e.\ when the cloud was magnetically dominated, the disk that
formed was much less prominent and did not fragment, but instead an outflow was driven
that removed of order 10\% of the gas that reached the disk, as in the \citet{mach06} study.

These results support the idea that outflows will be a natural consequence of the generation
of strong magnetic fields during Population III star formation. However, it is important to note
that the \citet{mach06,mach08} simulations only model the very earliest stages of outflow
driving, on a timescale $t \ll 1 \: {\rm yr}$. The evolution of outflows on much longer timescales,
and their influence on the infalling gas have not yet been studied in detail, and it is unclear
to what extent one can safely extrapolate from the very limited period that has been studied.

A strong magnetic field could also have a direct impact on the thermal evolution of the gas, through the
heating arising from ambipolar diffusion. The effects of this process in gravitationally collapsing gas
within the first star-forming minihalos have been investigated by \citet{sch09} using a simple one-zone
treatment of the gas. They assume that in the absence of ambipolar diffusion, the magnetic field
strength would evolve as $B \propto \rho^{\alpha}$, where $\alpha = 0.57 (M_{\rm J} / M_{\rm J, mag})^{0.0116}$
and $M_{\rm J, mag}$ is the magnetic Jeans mass (i.e.\ the minimum mass that a perturbation must have \index{Jeans mass!magnetic}
in order to be unstable to its own self-gravity when support against collapse is provided by a magnetic
field, rather than by thermal pressure). This expression for $\alpha$ is an empirical fitting formula
derived by \citet{sch09} from the results of \citet{mach06}. In their treatment of the evolution of $B$
within their one-zone models, \citet{sch09} also account for the loss of magnetic energy through
ambipolar diffusion.

Another important simplification made in the \citet{sch09} model is the replacement of the full expression for
the ambipolar diffusion heating rate \citep{pinto08}
\begin{equation}
L_{\rm AD} = \frac{\eta_{\rm AD}}{4\pi} \frac{| (\nabla \times {\mathbf B}) \times {\mathbf B}|^{2}}{B^{2}},
\end{equation}
where $B = |{\mathbf B}|$ and $\eta_{\rm AD}$ is the ambipolar diffusion resistivity, with the simpler approximation
\begin{equation}
L_{\rm AD} =  \frac{\eta_{\rm AD}}{4\pi} \frac{B^{2}}{L_{B}}, 
\end{equation}
where $L_{B}$ is the coherence length of the magnetic field.

\citet{sch09} show that if the initial field strength is less than 0.01~nG (comoving), then ambipolar diffusion 
heating has almost no effect on the thermal evolution of the gas. For stronger fields, ambipolar diffusion
heating leads to an increase in the gas temperature at densities between $n \sim 10^{4} \: {\rm cm^{-3}}$
and $n \sim 10^{10} \: {\rm cm^{-3}}$, amounting to as much as a factor of three at $n \sim 10^{8} \: 
{\rm cm^{-3}}$. However, at higher densities, three-body H$_{2}$ formation heating becomes a more
important heat source than ambipolar diffusion, meaning that the temperature evolution becomes largely
independent of the magnetic field strength once again. \citet{sch09} did not examine initial field strengths
larger than 1 comoving nG, as these are ruled out by observational constraints, but if one considers the
effects of the turbulent dynamo acting during the collapse, then it is possible that much larger fields could
be generated on smaller scales, and it would be useful to revisit this issue and examine whether 
ambipolar diffusion heating from these smaller-scale fields can significantly affect the collapse of the gas.

Finally, one important caveat to bear in mind regarding the \citet{mach06,mach08} and \citet{sch09}
simulations is the fact that they adopt a correlated initial magnetic field, while the field generated by
the turbulent dynamo will have little or no correlation on large scales \citep{maron04}. 
The extent to which the dynamical and thermal effects of this uncorrelated field will be the same as those 
of a correlated field is unclear. Further investigation of this issue would be very valuable.
\index{magnetic fields!effects|)}
\index{magnetic fields|)}
\index{protostar!formation|)}

\section{Evolution after the formation of the first protostar}
\label{later}
As we saw in the last section, when it comes to the formation of the
very first Population III protostar, there is broad agreement on the
details of the process, with different groups, who use different
numerical approaches, finding results that are in good qualitative
agreement with each other. Some quantitative disagreements still exist
\citep[see e.g.][]{turk11}, but it is unclear to what extent these 
reflect real differences between numerical approaches as opposed to
natural variation in the details of the collapse. The main uncertainties
in this phase stem from uncertainties in the input physics, such as 
whether magnetic fields can become amplified to dynamically significant
levels during the collapse, or whether dark matter annihilation 
significantly affects the outcome.

Once we move on to considering the evolution of the gas within
star-forming minihalos {\em after} the formation of the first protostar,
the situation becomes much less clear. The fundamental problem stems 
from the fact that although we can follow the gravitational collapse of
the primordial gas down to scales as small as the protostellar radius
\citep[see e.g.][]{yoh08}, the numerical timestep in an explicit hydrodynamical code 
becomes extremely short during this process. This is a consequence of
the Courant condition, which states that for such a code to be numerically
stable, the timestep must satisfy
\begin{equation}
\Delta t \leq \frac{\Delta x}{c_{s}},
\end{equation}
where $\Delta x$ is the size of the smallest resolution element, and $c_{s}$
is the sound speed of the gas.

The Courant condition implies that if we take a value of $\Delta x$ small
enough to adequately resolve the structure of the protostar and the gas
immediately surrounding it (e.g.\ $\Delta x = 1 {\rm R_{\odot}}$), then 
the required timestep will be extremely small: $\Delta t \leq 7 \times 10^{4} \: {\rm s}$
for $\Delta x = 1 {\rm R_{\odot}}$ and a sound speed of $10 \: 
{\rm km \: s^{-1}}$. This means that if we want to follow the later
evolution of the protostar and the surrounding gas over a timescale of 
thousands of years in order to see how it grows in mass prior to reaching
the main sequence, then we must use a very large number of timesteps: our
simple estimate above yields a number of the order of a million. In practice, the
computational expense of doing this within a three-dimensional hydrodynamical
code is prohibitively large, meaning that it has so far proved impossible to
study the evolution of the gas in this fashion.

Efforts to surmount this difficulty typically follow one of two approaches.
One approach is simply to halt the numerical simulation at the point at 
which the Courant timestep becomes prohibitively small, and to model the
later evolution of the protostar using a semi-analytical, or one-dimensional,
fully numerical treatment. To do this, it is necessary to make some assumption
about the behaviour of the gas surrounding the protostar. In general, models
of this type assume that the gas does not fragment and form additional 
protostars, but instead is simply accreted by the existing protostar, either
directly or via a protostellar accretion disk. The results obtained using this
approach -- what we afterwards refer to as the ``smooth accretion model''
-- are discussed in Section~\ref{mono} below.

The other approach that can be used to study the further evolution of the  
gas surrounding the protostar makes use of a technique developed for studies
of contemporary star formation, which face a similar problem on protostellar
scales. Gravitationally bound regions of gas that become smaller than some 
pre-selected size scale are replaced by what are usually termed sink 
particles \citep[see e.g.][]{bbp95}. These particles can accrete gas from  \index{sink particles|(}
their surroundings and continue to interact gravitationally with the 
surrounding gas, but allow one to neglect the very small-scale hydrodynamical
flows that would otherwise force one to take very small numerical timesteps
owing to the Courant condition. The great advantage of the sink particle
technique is that one need make no assumption about the dynamical evolution
of the gas surrounding the protostar on scales much larger than the effective
size of the sink particle (the so-called accretion radius, discussed in more
detail below), as one can simply continue to model this using the same 
numerical techniques as were used to model the initial gravitational collapse.
The main disadvantage of the technique is that, strictly speaking, it represents 
an {\em ad hoc} modification of the fluid equations, with consequences that
may not be entirely straightforward to predict. The modification to the solution
caused by replacing dense gas with sink particles is unlikely to significantly
affect the evolution of the gas on scales that are much larger than the accretion radius, 
but will clearly have an effect
on the flow on scales close to the accretion radius. In addition, the common
strategy of treating sink particles as point masses may not be appropriate when
dealing with close encounters between sinks, as one misses the tidal forces
acting between the gas clumps represented by the sinks. 

Although sink particles have been used in studies of Population III star formation
for over a decade, simulations using the correct initial conditions, and with sufficient 
spatial resolution and mass resolution to capture the details of the gas flow on scales 
close to those of individual protostars have only recently become possible. These
simulations show that, contrary to the assumption made in the smooth accretion
model, the gas generally fragments, rather than simply accreting onto a single,
central protostar. The results obtained from studies using sink particles --
afterwards referred to as the ``fragmentation model'' for Population III star formation 
-- are discussed in  Section~\ref{fragment} below.  \index{sink particles|)}

\subsection{The smooth accretion model}
\label{mono}

\subsubsection{Determining the accretion rate}
\index{protostar!accretion rate|(}
As we have already discussed above, at the point at which the protostar forms, its mass is
very small ($M \sim 0.01 \: {\rm M_{\odot}}$; see \citealt{yoh08}), 
but it is surrounded by an infalling envelope of gas containing tens or 
hundreds of solar masses. If we assume that the gas in this infalling envelope does not
undergo gravitational fragmentation, then it has only two possible fates -- it must either
be accreted by the central protostar (or protostellar binary; see e.g.\ \citealt{tao09}), 
or it must be prevented from
accreting, and possibly expelled from the immediate vicinity of the protostar, by some
form of protostellar feedback. This means that the mass of the protostar at the point at
which it forms has very little to do with its final mass. To determine the size of the latter,
we must understand the rate at which gas is accreted by the protostar, and how this
process is affected by protostellar feedback.

Since protostellar feedback involves a number of different processes, many of which are complicated
to model, it is easiest to start by considering models in which feedback effects are not included. As
feedback acts to reduce the accretion rate, models of this type allow us to place an upper limit on the
final mass of the Pop.\ III star.

A useful starting point is a simple dimensional analysis. Suppose that the protostar is embedded in
a gravitationally unstable cloud of mass $M$ and mean density $\langle \rho \rangle$, and that the
protostellar mass $M_{*} \ll M$, so that its gravity is negligible in comparison to the self-gravity of
the cloud. The timescale on which the gas cloud will undergo gravitational collapse and be accreted
by the protostar is simply the free-fall collapse time, $t_{\rm ff} = \sqrt{3\pi / 32 G \langle \rho \rangle}$.
Therefore, the time-averaged accretion rate will be given approximately by
\begin{equation}
\dot{M}_{\rm est}  \sim  \frac{M}{t_{\rm ff}} \sim M \sqrt{G \langle \rho \rangle}.
\end{equation}
If the gas cloud were highly gravitationally unstable, then it would fragment rather than accreting onto
a single object, so let us assume that it is only marginally unstable, i.e.\ that $M \sim M_{\rm J}$. In that
case, since $M_{\rm J} \sim c_{s}^{3} G^{-3/2} \rho^{-1/2}$, we can write our estimate of the time-averaged
accretion rate as
\begin{eqnarray}
\dot{M}_{\rm est}  & \sim &  M_{\rm J} \sqrt{G \langle \rho \rangle}, \\
 & \sim & \frac{c_{s}^{3}}{G}.
\end{eqnarray}
We therefore find that the characteristic accretion rate scales as the cube of the sound speed. Moreover,
since $c_{s} \propto T^{1/2}$, this implies that the accretion rate scales with temperature as $\dot{M}
\propto T^{3/2}$.

This is an important result, because as we have already seen, the characteristic temperature of the
dense, star-forming gas in a primordial minihalo is of the order of 1000~K, far larger than the 10~K
temperatures found within prestellar cores in local regions of star formation \citep[see e.g.][]{bt07}. 
Our simple scaling argument therefore tells us that we will be dealing with far higher accretion rates in the
Population III case than we are used to from studies of local star formation.

If we want to improve on this simple scaling argument and derive a more accurate figure for the
accretion rate, then there are three main ways in which we can go about it. One possible approach
is to construct a simplified model for the collapsing protostellar core from which an approximation
to the true accretion rate can be derived analytically (or with only minor use of numerical
calculations). For example, if we assume that the protostellar core is isothermal and spherically
symmetric, then there is a whole family of similarity solutions that could potentially be used to describe 
the collapse \citep{hunt77,ws85}, including the familiar Larson-Penston solution \citep{lar69,pen69}, or
the Shu solution \citep{shu77}.

An example of this approach is given in \citet{on98}. These authors used a spherically symmetric
Lagrangian hydrodynamical code to simulate the formation of a Population III protostar, and found
that prior to core formation, the gravitational collapse of the gas could be well described with
a Larson-Penston similarity solution, with an entropy parameter $K = p / \rho^{\gamma}
= 4.2 \times 10^{11}$ (in cgs units)
and an effective adiabatic index $\gamma_{\rm eff} = 1.09$. \citet{on98} were unable to continue their
numerical study past the point at which the protostar formed, for the reasons addressed above, but
assumed that the same similarity solution would continue to apply. By making this assumption, they
were therefore able to derive the following accretion rate for the protostar
\begin{equation}
 \dot{M} = 8.3 \times 10^{-2} \left( \frac{t}{\rm 1 \: yr} \right)^{-0.27}
\: {\rm M_{\odot}} \: {\rm yr}^{-1}.
\end{equation}
In a similar study, using a more sophisticated treatment of the microphysics of the collapsing
gas, \citet{ripa02} also found that the initial flow was well described as a Larson-Penston similarity 
solution, but derived a different accretion rate
\begin{equation}
 \dot{M} = 6.0 \times 10^{-2} \left( \frac{t}{\rm 1 yr} \right)^{-0.343}
 \: {\rm M_{\odot}} \: {\rm yr}^{-1}.
\end{equation}

Another example of this approach comes from \citet{tm04}. They model the accretion flow onto
a Pop.\ III protostar as a spherical, isentropic polytrope, and derive an accretion rate that is a 
function of three parameters: the entropy parameter $K$, the polytropic
index $\gamma_{p}$ (which, for an isentropic flow, is equal to the adiabatic  index $\gamma$), 
and $\phi_{*}$, a numeric parameter of order unity, which is related to the initial conditions of the 
flow. \citet{tm04} use the results of \citet{on98} and \citet{ripa02} to argue that $\gamma_{p} = 1.1$,
and use the 3D simulation results of \citet{abn02} to set the other two parameters to $\phi_{*} = 1.43$
and  $K = 1.88 \times 10^{12} K^{\prime}$ (in cgs units), where 
\begin{equation}
 K^{\prime} = \left(\frac{T_{\rm eff}}{300 \: {\rm K}}\right) 
 \left( \frac{n_{\rm H}}{10^{4} \: {\rm cm}^{-3}} \right)^{-0.1}, 
 \label{kprime}
\end{equation}
and where the effective temperature $T_{\rm eff} = P_{\rm eff} / (nk)$ accounts for the contribution
made to the pressure by small-scale, subsonic turbulence in addition to the standard thermal
pressure. Based on this, they then derive the following rate for the accretion of gas onto the
protostar and its associated accretion disk
\begin{equation}
 \dot{M} = 7.0 \times 10^{-2} {K^{\prime}\,}^{3/2} 
 \left( \frac{t}{\rm 1 \: yr} \right)^{-0.30} \: {\rm M_{\odot}} \: {\rm yr}^{-1}.   \label{tmrate}
\end{equation}
This can be directly compared to the other determinations of  $\dot{M}$ if we assume that all
of the gas reaching the accretion disk is eventually accreted by the star, which is a reasonable
assumption for models that do not include the effects of gravitational fragmentation or protostellar 
feedback.

Instead of using simulation results to select a particular collapse model (e.g.\ Larson-Penston
collapse) and then calculating $\dot{M}$ from the model, the second main approach used to 
determine $\dot{M}$ attempts to infer it from the state of the gas in the simulation at the point
at which the protostar forms, using the information that the simulation provides on the density
and velocity distributions of the gas. This approach was pioneered by \citet{abn02},
who considered two simple models for the time taken for a given fluid element to accrete onto
the central protostar. In the first of these models, they assumed that the time taken
for the gas within a spherically-averaged shell of radius $r$ to accrete onto the protostar was
given by the ratio between the mass enclosed within the shell, $M(r)$, and the rate at which
gas was flowing inward at that radius, i.e.\
\begin{equation}
t_{\rm acc} = \frac{M(r)}{4 \pi r^{2} \rho(r) |v_{\rm r}(r)|},
\end{equation}
where $\rho(r)$ and $v_{\rm r}(r)$ are the spherically-averaged density and radial velocity 
in the shell. In the second model for $t_{\rm acc}$, they used an even simpler
approximation, setting $t_{\rm acc}$ to the time that it would take for the gas to reach
the protostar if it merely maintained its current radial velocity, i.e.\
\begin{equation}
t_{\rm acc} = \frac{r}{v_{\rm r}}.
\end{equation}
\citet{abn02} show that other than at the very earliest times, these two approaches yield very
similar values for $t_{\rm acc}$, and hence very similar values for the accretion rate.
This strategy has subsequently been used by many other authors to derive predicted protostellar
accretion rates from their simulations \citep[see e.g.][]{yoha06,on07,mb08,turk11}. Of particular
note is the study by \citet{on07}, who perform multiple simulations of Population III star formation
using different random realizations of the cosmological density field. They find that minihalos
assembling at higher redshifts form more H$_{2}$ than those assembling at lower redshifts, 
owing to the higher mean density of the virialized gas in the high redshift minihalos. They show
that in their simulations, this leads to the gas at densities $n > 10^{4} \: {\rm cm^{-3}}$ having
significant differences in its mean temperature in the different halos. In the most H$_{2}$-rich
minihalos, the dense gas can be as cold as 200~K, while in the minihalos with the least
H$_{2}$, it can be as high as 1000~K. As a result, the predicted accretion rates for the different
minihalos span more than an order of magnitude, thanks to the strong scaling of $\dot{M}$
with temperature. Unfortunately, it is necessary to treat these results with a degree of caution,
as the \citet{on07} simulations did not include the effect of three-body H$_{2}$ formation heating,
which is known to have a significant influence on the temperature of the dense gas. It is unclear
whether simulations that include this effect would produce dense gas with such a wide range of
temperatures and accretion rates, although a study that is currently being carried out by
Turk and collaborators should address this issue in the near future (M.\ Turk, private communication).

The third main approach used to determine the accretion rate involves measuring it directly
in a simulation of the later evolution of the gas around the protostar. If we replace the
protostar with a sink particle, then we can measure $\dot{M}$ simply
by measuring the rate at which the sink particle mass increases. This approach was first used
by \citet{bl04}, in a study of Population III star formation in which a sink particle was created
once the gas density exceeded a threshold value $n_{\rm th} = 10^{12} \: {\rm cm}^{-3}$
(we will have more to say about this study below). \citet{bl04} showed that the rate at which
gas was accreted by the sink particle could be approximated as a broken power-law
\begin{equation}
 \dot{M} = \left\{ \begin{array}{lr}
 5.6 \times 10^{-2} \left(\frac{t}{1 {\rm yr}} \right)^{-0.25} \: {\rm M_{\odot}} \: {\rm yr^{-1}} &
 \hspace{.5in} t \leq 10^{3} \: {\rm yr} \\
 & \\
 6.3 \times 10^{-1} \left(\frac{t}{1 {\rm yr}} \right)^{-0.6}  \: {\rm M_{\odot}} \: {\rm yr^{-1}} &
 \hspace{.5in} t > 10^{3} \: {\rm yr}
                   \end{array} \right.   \label{blrate}
\end{equation}
for times $t < 10^{4} \: {\rm yr}$. Bromm \& Loeb halted their simulation at $t \sim 10^{4} \: {\rm yr}$ 
and hence could not directly measure the evolution of $\dot{M}$ at later times, although they
did consider what the final mass of the protostar would be if one simply extrapolated 
Equation~\ref{blrate} over the three million year lifetime of a massive star.

Accretion rates have also been measured using the sink particle technique in the group of 
simulations carried out by \citet{clark11a,clark11b}, \citet{greif11a} and \citet{smith11} 
that find evidence for fragmentation of the gas (see Section~\ref{fragment} below). The
accretion rates onto the individual sinks show a considerable degree of variability in these
calculations, but the {\em total} accretion rate, i.e.\ the rate of change of the sum of all of the
sink particle masses, evolves more smoothly with time, and is of a similar order of magnitude
to the other estimates plotted above.  

\begin{figure}
\includegraphics[scale=.45,angle=270]{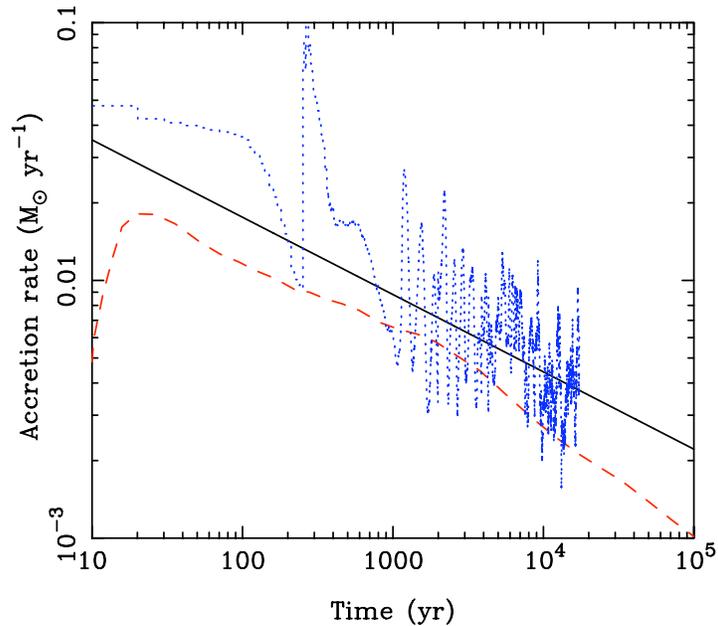}
\caption{Three different estimates for the accretion rate onto a Pop.\ III protostar, taken from
Tan \& McKee (2004; solid line), Turk et~al.\ (2011; dashed line) and Smith et al.\ (2011; dotted line), 
as described in the text. Results from the \citet{smith11} simulation are only plotted for the period 
covered by the simulation, i.e.\ $t < 2 \times 10^{4} \: {\rm yr}$.}
\label{fig:accrete} 
\end{figure}

In Figure~\ref{fig:accrete} we compare several of these different estimates for the accretion rate. 
We plot three examples, derived using different techniques: a rate based on the \citet{tm04}
formalism, computed assuming that $K^{\prime} = 1$;  a rate inferred from the results of one of the 
adaptive mesh refinement simulations presented in \citet{turk11} -- specifically, the simulation that 
was run using the \citet{pss83} rate coefficient for three-body H$_{2}$ formation; and a rate
measured using sink particles, taken from \citet{smith11}. 

At very early times ($t < 100 \: {\rm yr}$), the three different techniques yield rather different estimates
for $\dot{M}$, but this is primarily a consequence of the limited resolution of the numerical simulations.
At later times, we see that both the \citet{tm04} formalism and the \citet{turk11} simulation predict a 
similar form for the accretion rate, but disagree by about a factor of two on the normalization, which
may simply indicate that our adopted value of $K^{\prime}$ is slightly too large. We also see the same
general trend in the \citet{smith11} results, but in this case there is considerable and rapid variation in 
$\dot{M}$ with time. This is a result of the fragmentation of the gas in this simulation, which produces
a set of sink particles that undergo chaotic N-body interactions (see Section~\ref{fragment} below). 
A similar effect is seen in simulations of protostellar accretion in present-day star-forming regions
\citep[see e.g.][]{swh11}.

Regardless of whether $\dot{M}$ varies smoothly or erratically with time, one fact that is clear from
Figure~\ref{fig:accrete} is that the protostellar accretion rate remains very large for a considerable
time. This implies that the total mass of gas that is converted to stars can become fairly large after
a relatively short time. For example, if we take the \citet{tm04} estimate with $K^{\prime} = 1$ as a
guide, then we find that the total mass in stars increases with time as:
\begin{equation}
M_{*} = 0.1
 \left( \frac{t}{\rm 1 \: yr} \right)^{0.70} \: {\rm M_{\odot}} \: {\rm yr}^{-1}.  \label{tot_mass}
\end{equation}
This means that after $5 \times 10^{4} \: {\rm yr}$ (the Kelvin-Helmholtz relaxation time for a $100 \:
{\rm M_{\odot}}$ star), we have $M_{*} \simeq 195 \: {\rm M_{\odot}}$, while after $2 \times 10^{6}
\: {\rm yr}$ (the typical lifetime for an O star), we have $M_{*} \simeq 2575 \: {\rm M_{\odot}}$. Therefore, 
if the gas does not fragment and protostellar feedback is ineffective, one is led to the prediction that 
the resulting Population III star will be extremely massive. In practice, the gas probably does fragment
(see Section~\ref{fragment} below), and protostellar feedback cannot be completely ignored, but
even so, we would expect to be able to form massive Population III stars relatively easily.

Finally, it should be noted that so far we have considered accretion only in the standard 
H$_{2}$-dominated case, i.e.\ in a minihalo with a minimum gas temperature of around
200~K. In minihalos that reach much lower temperatures through HD cooling, the predicted
accretion rates are smaller, as one would expect from the simple dimensional analysis
given at the start of this section. For example, if one uses the \citet{tm04} formalism to
estimate the accretion rate, then Equation~\ref{tmrate} still applies, but the value of 
$K^{\prime}$ is significantly smaller. Taking $n = 10^{6} \: {\rm cm^{-3}}$ and
$T_{\rm eff} = 150$~K as plausible  values to substitute into Equation~\ref{kprime}, we
find that $K^{\prime} \simeq 0.3$, and hence the predicted accretion rate is roughly a factor
of six smaller than in the H$_{2}$-dominated case. Values estimated from numerical simulations
using the \citet{abn02} approach agree fairly well with this simple estimate 
\citep[see e.g.][]{yoh07,mb08}.
\index{protostar!accretion rate|)}

\subsubsection{Protostellar structure and evolution}
\label{proto}
\index{protostar!structure|(}
Having established how quickly gas will be accreted by the protostar in the absence of feedback,
the obvious next step is to examine how this will be modified by protostellar feedback. Before
doing this, however, we must first spend a little time discussing what is known about the internal
structure of Population III protostars, and how this evolves with time. This is important if we want
to understand how the radius and luminosity of a given Pop.\ III protostar evolve, and these
quantities are obviously of great importance when determining the influence of that protostar on 
the surrounding gas.

The internal structure of a Pop.\ III protostar, and how this evolves as the protostar ages
and accretes matter from its surroundings was first studied in detail by \citet{sps86a,sps86b}.
They assume that the accretion process can be treated as a series of quasi-steady-state 
accretion flows onto a hydrostatic core, which is bounded by a strongly radiating accretion 
shock. Within the core, the standard stellar structure equations are solved.
Outside of the core, the treatment depends on the optical depth of the gas. If
the gas is optically thin to the radiation from the accretion shock, then
the accretion flow is assumed to be in free-fall. Otherwise, a more detailed
calculation is made that incorporates the effects of the radiation force on
the infalling gas. The accretion shock itself is treated as a simple
discontinuity. 

In their initial study, \citet{sps86a} began with a core mass of $0.01\: {\rm M_{\odot}}$
and followed the growth of the protostar until its mass reached $10.5 \: {\rm M_{\odot}}$.
They assumed a constant accretion rate $\dot{M} = 4.41 \times 10^{-3} \: {\rm M_{\odot}} 
\: {\rm yr}^{-1}$, and found that for this choice of accretion rate, the evolution of the
protostar could be divided into three qualitatively distinct phases.

In the first phase, which lasts until the protostellar mass $M_{*} = 0.1 \: {\rm M_{\odot}}$,
the protostar relaxes from its initial entropy profile into one consistent with the selected
accretion rate. \citet{sps86a} dub this a `decay of transients' phase,   
and the fact that it quickly comes to an end shows that although the initial conditions
used in the \citet{sps86a} study are probably incorrect in detail, the flow soon loses all memory 
of them, and therefore any inaccuracy at this stage is unlikely to affect the later results.

Once the initial transients have died away, the protostar enters the second
phase of its evolution. During this phase, its central temperature remains
low ($T_{\rm c} \sim 10^{5} \: {\rm K}$), resulting in a high interior
opacity and hence a low interior luminosity. Consequently, the evolution of
the core during this phase is almost adiabatic; although the core continues
to gradually contract, this contraction does not lead to any increase in the
central entropy. Since the postshock entropy increases over time due to the
increasing strength of the accretion shock (which is itself a natural result
of the increasing protostellar mass), the core develops an off-centre
distribution of entropy and temperature.

The gas surrounding the accretion shock remains optically thick throughout
this period. This is a direct result of the high accretion rate, which
produces a highly luminous accretion shock. This produces sufficient
radiation to partially ionize the preshock gas in the vicinity of the
shock, creating a structure known as a radiative precursor. The H$^{-}$ opacity
of the dense, partially ionized gas in this radiative precursor is more than
sufficient to make it optically thick. \citeauthor{sps86a} show that the core
radius during this period evolves as
\begin{equation}
 R_{*} = 48.1 \left(\frac{M_{*}}{{\rm M_{\odot}}}\right)^{0.27}
 \left(\frac{\dot{M}}{\dot{M}_{0}}\right)^{0.41} \: {\rm R_{\odot}},
\end{equation}
where $\dot{M}_{0} = 4.41 \times 10^{-3} \: {\rm M_{\odot}} \: {\rm yr}^{-1}$, while
the photospheric radius evolves as
\begin{equation}
 R_{\rm p} = 66.8 \left(\frac{M_{*}}{{\rm M_{\odot}}}\right)^{0.27}
 \left(\frac{\dot{M}}{\dot{M}_{0}}\right)^{0.41} \: {\rm R_{\odot}},
\end{equation}
so $R_{\rm p} > R_{*}$ throughout. The strong H$^{-}$ opacity also keeps the
photospheric temperature low ($T_{\rm p} \sim 6000 \: {\rm K}$), which
prevents the protostar from being able to ionize material outside of its
photosphere.

This near-adiabatic accretion phase comes to an end once the cooling time of  
the core, given approximately by the Kelvin-Helmholtz timescale  \index{Kelvin-Helmholtz time}
\begin{equation}
 t_{\rm KH} = \frac{G M_{*}^{2}}{R_{*} L_{*}},
\end{equation}
becomes comparable to the accretion timescale $t_{\rm acc} = M_{*} / \dot{M}$.
This occurs for a core mass $M \sim 1 {\rm M_{\odot}}$, and results in the core
entering a phase of homologous collapse, while energy and entropy are
transferred outwards in the form of a `luminosity wave'. The radial position
of the luminosity peak moves outwards towards the accretion shock,
reaching it at about the time that the core mass has reached $8 \: {\rm M_{\odot}}$.
This results in a rapid swelling of the outermost layers, which weakens the
accretion shock and leads to it becoming optically thin. \citeauthor{sps86a}
terminate their simulation shortly afterwards, once the core mass has
reached $10.5 \: {\rm M_{\odot}}$.

\citet{sps86b} simulate the later stages of the evolution of a primordial protostar.
Their initial protostellar core has a mass of $5 \: {\rm M_{\odot}}$, and they
evolve this core forward in time, assuming that no further accretion occurs (i.e.\
the protostellar mass remains fixed at $5 \: {\rm M_{\odot}}$). They find that
deuterium burning within the protostar begins after only 6000~years, but that
hydrogen ignition does not occur until $t = 2 \times 10^{5} \: {\rm yr}$, and the
protostar does not reach the zero-age main sequence (ZAMS) until
$t \sim 10^{6} \: {\rm yr}$.

An improved treatment of the later stages of the evolution of the protostar was
made by \citet{op01}. They used a very similar setup to that in \citet{sps86a},
albeit with improved zero metallicity opacities, and adopted the same constant
accretion rate, $\dot{M} = 4.41 \times 10^{-3} \: {\rm M_{\odot}} \: {\rm yr}^{-1}$.
However, unlike \citet{sps86a}, they initialized their simulation at the point 
at which the core mass was  $M = 8 \: {\rm M_{\odot}}$, but did not halt the
simulation once the core had grown to $10.5 \: {\rm M_{\odot}}$. Instead, they
continued to follow the growth of the protostar until well after hydrogen ignition.
They found that deuterium burning within the core began once the core mass
was $12 \: {\rm M_{\odot}}$ (corresponding to a time $t = 1000 \: {\rm yr}$ after the
beginning of the simulation, given the assumed accretion rate), and that it was
complete by the time the mass had reached $30 \: {\rm M_{\odot}}$ (corresponding 
to $t = 5000 \: {\rm yr}$). Hydrogen ignition followed roughly 11000 years later,
at  $t = 1.6 \times 10^{4} \: {\rm yr}$ after the beginning of the simulation, at which
time the mass of the protostar was $80 \: {\rm M_{\odot}}$. At this point, the internal
luminosity of the protostar is very close to the Eddington value, which leads to the
outer layers of the protostar developing oscillatory behaviour: the high luminosity
leads to expansion, the expansion causes the accretion luminosity to drop, the
reduced luminosity can no longer maintain the expansion, leading to contraction
of the core, and the contraction raises the accretion luminosity, allowing the whole
cycle to begin again.  Finally, once the core mass reaches $300 \: {\rm M_{\odot}}$,
at $t \sim 6.6 \times 10^{4} \: {\rm yr}$, the contribution of nuclear burning to the protostellar 
luminosity becomes large enough to drive a final phase of expansion that is strong
enough to terminate accretion onto the protostar. \citet{op01} halt their simulation
at this point. 

In a follow-up study using a similar spherically-symmetric setup, \citet{op03}
performed the same analysis for a range of different values of $\dot{M}$, 
looking at models with $\dot{M} = (0.25, 0.5, 1.0, 2.0) \times \dot{M}_{\rm fid}$
(where $\dot{M}_{\rm fid}$ was the rate adopted by \citealt{sps86a} and \citealt{op01}),
as well as a model using the time-dependent accretion rate predicted by \citet{abn02}.
The earliest stages of protostellar evolution are qualitatively the same in
all of these models: we see again the same sequence of adiabatic growth,
propagation of a luminosity wave that triggers expansion of the outer layers,
and then rapid contraction. Although some quantitative differences are apparent,
significant differences in behaviour do not occur until the end of the contraction phase.
At this point, the further evolution of the protostar is governed by the size of the
accretion rate. For accretion rates greater than some critical value $\dot{M}_{\rm crit}$, the luminosity of the protostar
becomes large enough to halt the accretion. On the other hand, for $\dot{M} < \dot{M}_{\rm crit}$,
the lower accretion luminosity means that the total luminosity of the protostar remains
below $L_{\rm Edd}$, and accretion continues unabated. 

\citet{op03} solve for $\dot{M}_{\rm crit}$ by equating the total luminosity of a zero-age main
sequence Pop.\ III protostar (including accretion luminosity) with the Eddington luminosity,
and find that
\begin{equation}
\dot{M}_{\rm crit} \simeq 4 \times 10^{-3} \: {\rm M_{\odot}} \: {\rm yr}^{-1},
\end{equation}
coincidentally close to $\dot{M}_{\rm fid}$. In principle, one would expect
$\dot{M}_{\rm crit}$ to have a dependence on the current mass of the protostar,
but in practice, \citet{op03} show that this dependence is weak and may be neglected.

Finally, \citet{op03} show that in the time-dependent accretion model, the key factor is
the size of the accretion rate at the end of the contraction phase. If this is greater than
$\dot{M}_{\rm crit}$, then one would expect accretion to be halted, while if it is less than
$\dot{M}_{\rm crit}$ then accretion can continue. In practice, \citet{op03} show that if one
adopts the \citet{abn02} estimated accretion rate, then $\dot{M} < \dot{M}_{\rm crit}$,
implying that accretion can continue even once the protostar reaches the zero-age
main sequence.

\index{accretion disk|(}
The main limitation of the approach outlined above is the 
neglect of the effects of rotation. In reality, rotation can have profound
effects on stellar structure and evolution, particularly for massive stars \citep{mm00},
and it will also have a large influence on how matter reaches the protostar in the
first place. The first detailed study of the pre-main sequence evolution of a Pop.\
III protostar to account for the effects of rotation was carried out by \citet{tm04}.
In contrast to previous authors, they did not assume spherical symmetry. Instead,
they assumed that a protostellar accretion disk would form, and fixed the size of the
disk by assuming angular momentum conservation within the supersonic portion
of the accretion flow. They used the polytropic accretion flow model described in the 
previous section to compute the accretion flow onto the disk. To solve for the disk structure, they
made use of the standard theory of steady, thin viscous accretion disks (as outlined in 
\citealt{ss73}), with a spatially constant viscosity parameter $\alpha$.
As sources for $\alpha$, they considered the magnetorotational instability \citet{bh91,bh98}
and gravitational instability. With the disk structure in hand, they could then solve
for the structure of the protostar itself, using a modified version of an approach
developed by \citet{nhn95} and \citet{nhmy00}. In the zero angular momentum case,
\citet{tm04}  show that they successfully reproduce the previous results of \citet{sps86a}
and \citet{op01,op03}. In more realistic models, \citet{tm04} show that the presence of
an accretion disk has little influence on the evolution of the protostar, which still evolves 
through the same progression of adiabatic growth, terminated by the emergence of a luminosity 
wave, followed by rapid contraction to the ZAMS. However, \citet{tm04} do find that the 
photosphere surrounding the protostar behaves very differently in this case than in the
spherical case.  Because most of the gas accretes onto the protostar via the disk, the gas
density is significantly reduced in the polar regions. 
Consequently, the optical depth of these regions is
also significantly reduced, with the result that the flow becomes optically
thin early in its evolution. For example, in the model with $f_{\rm Kep} = 0.5$, the photosphere 
vanishes once the protostellar mass reaches $1 \: {\rm M_{\odot}}$ and does not subsequently 
reappear. \citet{tm04} argue that this may have a major influence on the effectiveness of radiative
feedback from the protostar, a topic that we will return to in the next section.
\index{protostar!structure|)}
\index{accretion disk|)}

\subsubsection{Feedback effects}
\label{feed}
Accretion of gas onto the protostar liberates a significant amount of energy, with most of this
energy being emitted from regions close to the protostellar surface. This can be shown very
simply by considering how the gravitational potential energy of a test mass changes as we
move it close to a protostar of mass $M_{*}$ and radius $R_{*}$. At a distance of $2R_{*}$,
the gravitational potential energy of a fluid element with mass $dM$ is
\begin{equation}
W = - \frac{GM_{*} dM}{2R_{*}},
\end{equation}
while at the protostellar surface it is
\begin{equation}
W = - \frac{GM_{*} dM}{R_{*}}.
\end{equation}
Therefore, the amount of energy that must be dissipated by the fluid element as it moves from
$2R_{*}$ to $R_{*}$ is as large as the amount that it must have dissipated while falling in from
$R \gg R_{*}$ to $2R_{*}$, or in other words, {\em half} of the total binding energy dissipated by
the gas is dissipated while its distance from the protostellar surface is less than $R_{*}$. In addition, 
once  the protostar reaches the main sequence, it will start generating additional energy in its own right, 
via nuclear fusion. The energy that is released in the vicinity of the protostar is therefore quite
considerable, and it is reasonable to suppose that this will have some effect on the behaviour
of the surrounding gas. It is therefore not surprising that considerable attention has been paid
to the issue of protostellar feedback in the context of Pop.\ III star formation.

In order for the protostar to substantially reduce the rate at which matter flows onto it, it must
be able to transfer a significant amount of energy and/or momentum to the infalling gas. The
various mechanisms by which this can be accomplished fall under two broad headings:
{\em mechanical feedback}, where the protostar transfers energy and momentum to some form of outflow, 
which subsequently transfers it to the infalling material, and {\em radiative feedback}, where
radiation from the protostar transfers energy and momentum directly to the infalling gas.

\paragraph{\bf Mechanical feedback}
\index{feedback!mechanical|(}
In the local Universe, stellar winds are an almost ubiquitous phenomenon, and play an important
role in the evolution of the most massive stars  \citep{cm86}. However, there are good reasons to
expect that metal-free stars will be much less effective at driving winds than the roughly solar metallicity 
stars that we are familiar with in the Milky Way. Strong stellar winds are invariably radiation-driven,
and at solar metallicities, the largest contribution to the radiative acceleration of the gas comes from
the absorption and scattering of ultraviolet photons in the lines of the many metal atoms and ions
present in the outflowing gas \citep{cak75}. In metal-free gas, on the other hand, the only
significant sources of opacity within an outflow will come from the lines of He$^{+}$ (atomic
hydrogen is typically fully ionized), and from Thomson scattering by free electrons. These
provide orders of magnitude less radiative acceleration per unit luminosity than do the metal
lines in a solar metallicity gas, and hence one can show that a metal-free Population III 
star can produce a line-driven wind only if the stellar luminosity is already very close to the Eddington
limit \citep{kud02}. 

Of course, as a Population III star evolves, it will not remain metal-free. It will start to produce 
carbon, nitrogen and oxygen internally once the stellar core begins to burn helium, and if the
star is rotating, these elements can become well-mixed within the star \citep{mem06}. This
will provide an additional source of opacity in the stellar atmosphere which may allow the most
massive Population III stars to produce a weak CNO-driven wind \citep{kk09}. However, the
mass-loss rate will be small, and the fraction of the stellar mass that  can lost in this way is
unlikely to be larger than about 1\%.

It is also possible that very massive Population III stars with luminosities close to the Eddington
luminosity may produce eruptive, continuum-driven winds, similar to those we see coming from 
nearby luminous blue variables (LBVs) such as $\eta$ Car \citep{so06}. However, as the cause of
these LBV eruptions is not yet fully understood even for nearby objects, it is difficult to say
with certainty whether they will actually be produced by Pop.\ III stars. More work on this topic
is clearly necessary.

Finally, mechanical feedback can also be generated in the form of hydrodynamical or
magnetohydrodynamical 
jets or outflows. We have already discussed the magnetically-driven disk winds produced in the  
\citet{mach06,mach08}  simulations, which are able to eject roughly 10\% of the  infalling gas from 
the disk. Although, as we noted previously, these simulations only modelled the very earliest stages
in the formation of the protostellar accretion disk, their value for the mass ejection rate is in good
agreement with the predictions of a semi-analytical study of Pop.\ III disks and outflows carried
out by \citet{tb04}. If this value is correct, then it implies that the reduction in the protostellar accretion
rate brought about by these outflows is small, and hence that they will not significantly limit the
final stellar mass. However, one should bear in mind that their interaction with the star-forming 
halo on larger scales has not yet been modelled in any detail, and hence it is difficult to be
certain regarding their final impact.
\index{feedback!mechanical|)}

\paragraph{\bf Radiative feedback}
\index{feedback!radiative|(}
There are several different forms of radiative feedback that could potentially affect the
accretion of gas by a Pop.\ III protostar. First, if the radiation is absorbed or scattered,
then it will exert a force on the gas. If this force is comparable to or larger than the
gravitational force acting on the gas, then it may suppress accretion onto the protostar,
or even prevent it completely. Second, radiation may destroy the H$_{2}$ molecules
responsible for cooling the gas. In the absence of cooling, the gas will evolve adiabatically,
which again may reduce the rate at which it can be accreted. Third, the radiation may
heat the gas. If radiative heating raises the gas temperature to a point at which the thermal
energy of the gas exceeds the gravitational binding energy of the system, then this again
will strongly suppress accretion.  

\index{radiation pressure|(}
In local star-forming regions, the first of these three forms of radiative feedback is believed
to be the most important. Radiation pressure exerted on infalling dust grains by radiation
from the protostar results in a substantial momentum transfer to the dust, and from there
to the gas, since the dust and gas are strongly coupled. In spherically symmetric models,
the radiative force exerted by the radiation on the dust can be strong enough to bring
accretion to a complete halt \citep{wc87}. In primordial gas, there is no dust, and so this 
process cannot operate.  However, radiation pressure can also work directly on the gas, 
and so it is worthwhile investigating whether this process is likely to significantly suppress 
accretion.

Let us start by assuming that the bolometric luminosity of the protostar is given by the Eddington 
luminosity \index{Eddington luminosity}
\begin{equation}
L_{\rm Edd} = \frac{4 \pi G M_{*} c}{\kappa_{T}},
\end{equation}
where $M_{*}$ is the protostellar mass, and $\kappa_{T} \equiv \sigma_{T} / m_{\rm p} \simeq
0.4 \: {\rm cm^{2}} \: {\rm g^{-1}}$ is the opacity 
due to Thomson scattering for a fully ionized gas composed of pure hydrogen, with $\sigma_{T}$ 
the Thomson scattering cross-section of the electron and $m_{\rm p}$ the mass of the proton. 
In this case, then we know from the definition of the Eddington luminosity that
the radiative force exerted on a fluid element will be equal to the gravitational force 
exerted on it by the protostar when the opacity of the fluid element is equal to $\kappa_{T}$.
More generally, we can write the ratio of the forces acting on the fluid element as
\begin{equation}
\frac{F_{\rm rad}}{F_{\rm grav}} = \frac{L_{*}}{L_{\rm Edd}} \frac{\kappa}{\kappa_{T}}, \label{mean_k}
\end{equation}
where $F_{\rm rad}$ is the radiative force, $F_{\rm grav}$ is the gravitational force, $L_{*}$
is the protostellar luminosity, and $\kappa$ is the mean opacity of the fluid element. 
Since the protostar is unlikely to be stable if $L_{*} > L_{\rm Edd}$, this implies that 
in order for the radiative force to significantly affect the gas, it must have a mean opacity
$\kappa \sim \kappa_{T} $ or higher. In practice, the luminosity of a Pop.\ III protostar 
before it reaches the main sequence will often be significantly less than the Eddington 
luminosity \citep[see e.g.][]{smith11}, in which case an even higher mean opacity is required.

The mean opacity of metal-free gas has been computed by a number of authors, most recently by
\citet{md05b}. They present tabulated values for both the Rosseland mean opacity  \index{Rosseland mean opacity}
\begin{equation}
\kappa_{R}^{-1} = \frac{\int_{0}^{\infty} (\partial B_{\nu} / \partial T) \kappa_{\nu}^{-1} \; d\nu}{\int_{0}^{\infty} 
(\partial B_{\nu} / \partial T) d\nu},
\end{equation}
and the Planck mean opacity  \index{Planck mean opacity|(}
\begin{equation}
\kappa_{P} = \frac{\int_{0}^{\infty} B_{\nu} \kappa_{\nu} \; d\nu}{\int_{0}^{\infty} B_{\nu} d\nu},
\end{equation}
where $\kappa_{\nu}$ is the frequency-dependent opacity and $B_{\nu}$ is the Planck function.
For our purposes, we are most interested in the Planck mean. Strictly speaking, this Planck mean 
opacity is the same as the mean opacity in Equation~\ref{mean_k} only if the protostar has a black-body radiation 
field and a photospheric temperature that is the same as the gas temperature, and in general this will not be
the case.  However, if the protostar is still in the pre-main sequence phase of its evolution, it will have a photospheric 
temperature $T_{\rm p} \sim 6000$~K \citep{sps86a} and a spectrum that does not differ too greatly from a black-body, 
while the temperature of the surrounding gas will typically be of the order of 1000-2000~K or higher
\citep{clark11b, greif11a,smith11}. In these conditions, the error we make by using the Planck mean opacity in 
Equation~\ref{mean_k} should not be excessively large. \index{Planck mean opacity|)}

There are two regimes in which the tabulated values of $\kappa_{P}$ in \citet{md05b} exceed $\kappa_{T}$.
The first occurs at very high densities ($n > 10^{22} \: {\rm cm^{-3}}$), where $\kappa_{P} > \kappa_{T}$ for a 
wide range of temperatures. However, these extreme densities are only reached {\em within} the protostar and
hence this regime is of no relevance when we are considering feedback from the protostar on the surrounding gas. 
The second regime in which $\kappa_{P}$ grows to the required size is at temperatures above 8000~K, for a wide
range of densities. At these temperatures, the dominant source of opacity is the scattering of photons in the Lyman
series lines of hydrogen, primarily Lyman-$\alpha$. The effects of Lyman-$\alpha$ radiation pressure in metal-free
gas were considered by \citet{oh02}, in the context of the formation of massive star-forming minihalos with
virial temperatures $T > 10^{4} \: {\rm K}$. They argued that the Lyman-$\alpha$ photons produced by the cooling
of the hot gas would not be important \citep[see also][]{ro77}, but that the Lyman-$\alpha$ photons produced by a 
massive star and its associated HII region would have a pronounced effect on the gas, and could significantly delay 
or even halt the inflow of the gas. However, they did not carry out a full quantitative investigation of the effects of 
Lyman-$\alpha$ radiation pressure. More recently, this issue was revisited by \citet{mt08}, who studied it in some
detail. They found that in a rotating flow, most of the Lyman-$\alpha$ photons would eventually escape along the
polar axis of the flow, as it is here that the optical depths are smallest. They showed that if the rotational speed of
the gas were at least 10\% of the Keplerian velocity, then Lyman-$\alpha$ radiation pressure would be able to
reverse the direction of the flow along the polar axis once the protostellar mass reached $20 \: {\rm M_{\odot}}$.
The radiation would therefore blow out a polar cavity, allowing more Lyman-$\alpha$ photons to escape.
This prevents the radiation pressure from rising further, and McKee \& Tan argue that it never becomes large
enough to significantly affect the inflow of gas from directions far away from the polar axis (e.g.\ from the 
accretion disk). For this reason, they conclude that Lyman-$\alpha$ radiation pressure is unlikely to be able to
significantly reduce the protostellar accretion rate.
\index{radiation pressure|)}

\index{photodissociation|(}
Let us now turn our attention to the second form of radiative feedback mentioned above: the photodissociation 
of H$_{2}$ and the consequent dramatic reduction in the cooling rate. As we have already discussed, at early
times the photospheric temperature of the protostar is too low for it to produce significant quantities of far-ultraviolet 
radiation, and hence radiation from the protostar does not significantly affect the H$_{2}$. Once the protostar reaches
the main sequence, however, it can become a significant source of far-ultraviolet radiation, provided that it has a
mass greater than around $15 \: {\rm M_{\odot}}$ \citep{mt08}.
Studies by \citet{on99} and \citet{gb01} considered the effect that this radiation would have on the 
H$_{2}$ surrounding the protostar, and showed that the time required to photodissociate the H$_{2}$ would be 
significantly less than the lifetime of the protostar. The removal of the H$_{2}$ from the gas means that it is no
longer able to cool effectively at temperatures $T < 10^{4} \: {\rm K}$, and hence one would expect that as the
H$_{2}$ in the accreting gas is destroyed, the gas will begin to evolve adiabatically until it reaches this temperature.
\citet{mt08} consider whether this switch to adiabatic evolution is sufficient to halt accretion, and conclude
that it is not.  If no protostar were present, then the switch to adiabatic evolution would be enough to stabilize
the gas and prevent further collapse. The presence of the protostar, however, serves to destabilize the gas,
allowing accretion to continue even when the evolution of the gas is fully adiabatic. \citet{mt08} use the
treatment of protostellar accretion introduced in \citet{fat04} to investigate the issue numerically, and show that
an increase in the effective adiabatic increase of the gas from $\gamma_{\rm eff} = 1.1$ (which approximately
characterizes the temperature evolution of the gas at $n > 10^{4} \: {\rm cm^{-3}}$; see e.g.\ \citealt{on98}) to 
$\gamma_{\rm eff} = 5/3$ reduces the accretion rate by only 20\%. 
\index{photodissociation|)}

The third possible form of radiative feedback involves the heating of the surrounding gas by radiation from 
the protostar. If the temperature of the gas can be increased to a point at which its thermal energy exceeds
its gravitational binding energy, then it will no longer be gravitationally bound to the protostar, and hence
will not be accreted. A convenient way to quantify the relative importance of thermal and gravitational
energy is to compare the sound-speed of the gas with the escape velocity of the system, $v_{\rm esc}$:
gas with $c_{\rm s} > v_{\rm esc}$  will not be gravitationally bound.

For an isolated protostar of mass $M_{*}$, we can write $v_{\rm esc}$ at a distance $R$ from the protostar as:
\begin{equation}
v_{\rm esc} = \sqrt{\frac{2 G M_{*}}{R}}, \label{vesc}
\end{equation}
where $G$ is the gravitational constant. If we rewrite this expression in more convenient units, we find that
\begin{equation}
v_{\rm esc} \simeq 4.2 \left(\frac{M_{*}}{1 \: {\rm M_{\odot}}}\right)^{1/2} \left(\frac{R}{100 \: {\rm AU}} \right)^{-1/2} \: {\rm km \: s^{-1}}.
\end{equation}
For a primordial, fully molecular gas, $c_{\rm s} = 4.2 \: {\rm km \: s^{-1}}$ at a temperature $T \sim 3400 \: {\rm K}$,
and hence gas within $100 \: {\rm AU}$ of a one solar mass protostar must be heated up to a temperature of
thousands of Kelvin in order to unbind it. At larger distances, the required temperature would appear at first to
be much smaller, but the reader should recall that this expression is for an {\em isolated} protostar, i.e.\ one which
is not surrounded by gas. It is therefore only valid when the protostellar mass $M_{*}$ is much larger than the
mass of  gas within a distance $R$ of the protostar, and once we start considering scales $R \gg 100 \: {\rm AU}$, 
this is unlikely to be a good approximation.  If we include the influence of this gas by replacing $M_{*}$ in 
Equation~\ref{vesc} by $M_{\rm tot} = M_{*} + M_{\rm gas}$, and use the facts that prior to star formation,
the mass enclosed within a sphere of radius 100~AU is roughly $5 \: {\rm M_{\odot}}$ and increases at
larger distances as $M_{\rm enc} \propto R^{0.8}$, then at distances $R > 100 \: {\rm AU}$, we have
\begin{equation}
v_{\rm esc} \simeq 9.4 \left(\frac{R}{100 \: {\rm AU}} \right)^{-0.1} \: {\rm km \: s^{-1}}.
\end{equation}
In other words, once we account for the mass of the infalling gas in addition to the mass of the protostar,
we find that the escape velocity is of the order of $10 \: {\rm km \: s^{-1}}$, with little dependence on the
distance from the protostar. An escape velocity of this order of magnitude corresponds to a gas temperature
of order $10^{4}$~K. This immediately tells us that heating of the gas by radiation from the protostar during
the pre-main sequence phase of its evolution is unlikely to significant affect the accretion rate due to the
low photospheric temperature of the protostar during this phase -- clearly, a protostar with an effective
temperature of $6000 \: {\rm K}$ will not be able to heat up distant gas to a temperature of 10000~K.
On the other hand, once the protostar reaches the main sequence, its photospheric temperature will
sharply increase, and hence it may be able to heat up the surrounding gas to a much higher temperature.
In particular, if the protostar is massive enough to emit a significant number of ionizing photons while
on the main sequence, then it will easily be able to produce temperatures in excess of $10^{4} \: {\rm K}$
within the gas that it ionizes.  

\index{photoionization|(}
The idea that the formation of an HII region may strongly suppress or completely terminate protostellar
accretion was discussed long ago in the context of present-day star formation \citep[see e.g.][]{ls71}, but 
has recently been re-examined by several authors in the context of primordial star formation. On large
scales ($R > 0.1 \: {\rm pc}$), the behaviour of an HII region produced by a Pop.\ III star is relatively simple. 
The radial density profile of the gas on these scales is approximately $\rho \propto R^{-2.2}$, and hence
the density falls off too quickly to trap the HII region within the minihalo \citep[see e.g][]{wan04,abs06,awb07,yokh07}. 
The ionization front therefore expands rapidly, as an R-type front, with a velocity that is controlled by the 
rate at which ionizing photons are being produced by the star. In addition, if we are considering Pop.\ III star 
formation within one of the first star-forming minihalos, then it is easy to show that sound speed of the gas 
within the HII  region will be higher than the escape velocity of the minihalo. Consequently, the ionized gas
begins to flow out of these small minihalos, significantly reducing the mean gas density. It is therefore clear
that once the HII region reaches a size of $0.1 \: {\rm pc}$ or above, it will act to prevent any further infall of
gas from these scales onto the protostar. However, this leaves unanswered the question of how long it
takes for the HII region to expand to this scale.

In the case of steady, spherically-symmetric infall, \citet{oi02} showed that in order for the HII region
to avoid being trapped on scales close to the protostar, the flux of ionizing photons must exceed a 
critical value
\begin{equation}
\dot{N}_{\rm crit} = 6.4 \times 10^{52} \left( \frac{R_{\rm in}}{10 \: {\rm R_{\odot}}}
\right)^{-1} \left( \frac{M_{*}}{100 \: {\rm M_{\odot}}} \right)^{2} \: {\rm s^{-1}},
\end{equation}
where $R_{\rm in}$ is the inner radius of the HII region, which we can take to be equal to the
radius of the massive star. Given reasonable values for $R_{\rm in}$ and $M_{*}$, this expression
yields a value for $\dot{N}_{\rm crit}$ that is much larger than the number of photons that will actually
be produced by any massive star, leading \citet{oi02} to conclude that the HII region would remain
trapped close to the star. However, this conclusion depends crucially on the assumed spherical
symmetry of the flow. In the more realistic case in which our protostar is surrounded by an accretion
disk, \citet{mt08} show that the HII region can expand in all directions other than those close to the
midplane of the disk once the stellar mass reaches a value of around 50--100$\: {\rm M_{\odot}}$,
where the precise value required depends on how rapidly the gas is rotating. \citet{mt08} also
show that the accretion disk can survive for a considerable period after the HII region has broken
out, and that the protostar will stop accreting from the disk only once the rate at which gas is
lost from the disk by photoevaporation exceeds the rate at which fresh gas is falling onto the
disk. In their models, this occurs for $M_{*} \sim 140 \: {\rm M_{\odot}}$, leading them to conclude
that radiative feedback from the protostar on the surrounding gas cannot prevent the protostellar
mass from becoming very large. On the other hand, initial attempts to model this process in 2D
or 3D \citep{hoso11,sgb12} find significantly smaller final masses, $M_{*} \sim 30$--$40 \:
{\rm M_{\odot}}$, although these detailed models have so far explored only a small part of the
potential parameter space.
\index{feedback!radiative|)}
\index{photoionization|)}

\subsection{The fragmentation model}
\label{fragment}
\index{fragmentation|(}
\subsubsection{Early studies}
The first simulations of primordial gas to make use of sink particles were the SPH simulations
of \citet{bcl99,bcl02}. They studied the formation of isolated dark matter minihalos and the
cooling and gravitational collapse of gas within them using a somewhat idealized set of initial
conditions. At an initial redshift $z=100$, they created a spherical, uniform density region
containing both gas and dark matter, and with an initial velocity field corresponding to the
Hubble expansion. The density of this spherical region was taken to be higher than the 
cosmological background density, and was fixed such that the region would gravitationally
collapse and virialize at a specified redshift, chosen to be $z = 30$ in most of the models that
they examined. Small-scale structure was introduced into the dark matter
distribution by perturbing the particles slightly from their initial positions using the 
\citet{zel70} approximation. The amplitudes of these random perturbations were fixed such
that the small-scale density structure in the dark matter would begin to evolve in the non-linear
regime at the virialization redshift. Both the dark matter and the gas were also assumed to be
in solid body rotation, with some specified angular velocity.

\citeauthor{bcl02} examined several different choices for the halo mass and initial angular
velocity of the gas, and showed that starting from these initial conditions, the gas and dark matter
would initially collapse in a similar fashion, but that the gas would subsequently form H$_{2}$,
dissipate energy, and sink to the center of the minihalo. In most of the cases they studied, the
gas would then form a rotationally-supported disk, with a radius of order 10~pc. This disk
would then break up into clumps with masses $M_{\rm cl} \sim 100$-1000~${\rm M_{\odot}}$,
comparable to the Jeans mass in the disk. As these clumps were gravitationally unstable, \index{Jeans mass}
they of course underwent gravitational collapse, and  \citet{bcl02} therefore introduced sink
particles to represent clumps that collapsed to densities greater than $10^{8} \: {\rm cm^{-3}}$
in order to avoid the timestep constraints discussed earlier, allowing the further evolution of
the clumps to be studied.

Unfortunately, the fragmentation observed by \citeauthor{bcl02} in their simulations is probably
not realistic. One major problem lies in their choice of initial conditions, specifically in their
use of solid-body rotation. Although the total angular momentum of the gas and dark matter
in their simulations is comparable that measured for minihalos in more realistic cosmological
simulations \citep{hjc01,dn10}, their adoption of solid-body rotation leads to the gas having an 
incorrect radial profile for this angular momentum. This causes the collapse of the gas to be 
considerably more ordered than it would be in a real minihalo, leading to the formation of an 
over-large disk. 
Disks of this kind do not appear to form in simulations of small star-forming minihalos that start 
from more realistic cosmological initial conditions \citep[e.g.][]{abn02,yoha06}. A second major 
problem lies in the neglect of stellar feedback. Within the disk, the dynamical timescale is of the 
order of a million years, which is much longer than is needed for a massive Pop.\ III star to reach 
the main sequence. Therefore, if a massive star forms within the first clump to be produced within 
the disk, the radiation from this star may well be able to photodissociate the H$_{2}$ in the disk
before a second clump can form \citep{on99,gb01}. 

The next attempt to use sink particles to study the formation of Pop.\ III stars was made by
\citet{bl04}. The initial setup of their simulation was similar to that used by \citet{bcl02}, but
to gain improved resolution in the centre of the minihalo, they used a technique called 
particle splitting \citep{kw02,bl03}. The evolution of the minihalo was followed until
cold, dense gas started to build up in the centre of the halo. The simulation was then
paused, and the gas within a radius of 3.1~pc of the centre of the minihalo (corresponding
to roughly $3000 \: {\rm M_{\odot}}$ of material) was resampled using SPH particles with
much smaller masses, using the resampling technique described in \citet{bl03}. The
mass resolution within this resampled region was thereby improved from $M_{\rm res}
= 200 \: {\rm M_{\odot}}$ to $M_{\rm res} = 4 \: {\rm M_{\odot}}$. Bromm \& Loeb then
restarted the simulation, and followed the further gravitational collapse of the gas within
this central, higher resolution region until the gas density reached $n \sim 10^{12} \:
{\rm cm^{-3}}$, at which point a sink particle was created. They then followed the accretion
of gas onto this sink for roughly $10^{4} \: {\rm years}$, as we have already described above.
\citet{bl04} found no evidence for fragmentation within the central clump of dense gas, but
did note that a second dense clump formed nearby, with a final separation from the 
star-forming clump of roughly 0.25~pc. However,  the free-fall collapse time of this clump
was about 3~Myr, and so it was unclear whether it would survive for long enough to form
a second star, or whether it would be destroyed by negative feedback from a massive star
forming within the first clump.

\subsubsection{The importance of turbulence}
Although the \citet{bl04} study undoubtedly represented a significant step forwards in
resolution compared to \citet{bcl02}, it still had a mass resolution which was more than
two orders of magnitude greater than the actual size of a Pop.\ III protostar at the moment
that it forms, and hence it was unable to investigate the behaviour of the gas on scales
smaller than about 100~AU. The first work using sink particles that did manage to probe
this regime was \citet{cgk08}. Although the main focus of their study was on the fragmentation
brought about by dust cooling in low metallicity systems (see e.g.\ \citealt{om05}, \citealt{sch06}
or \citealt{dopcke11} for more on this topic), they also studied the behaviour of the 
gas in the ${\rm Z} = 0$ case for the purpose of comparing it with the results on their low 
metallicity runs. As the initial conditions for their simulations, \citet{cgk08} considered a
uniform density cloud, with a mass of $500 \: {\rm M_{\odot}}$, a radius of 0.17~pc and
a number density of $5 \times 10^{5} \: {\rm cm^{-3}}$. The gas within this cloud was given
a low level of initial turbulence, with a turbulent energy equal to 10\% of the gravitational
potential energy, and was also assumed to be rotating uniformly, with an initial rotational
energy equal to 2\% of the gravitational potential energy. Two different simulations were
performed, with different numbers of particles: a low resolution simulation that used only
two million SPH particles, and hence had a mass resolution of $0.025 \: {\rm M_{\odot}}$,
and a high resolution simulation that used 25 million particles, corresponding to a mass
resolution of $2 \times 10^{-3} \: {\rm M_{\odot}}$. 
Aside from the somewhat artificial initial conditions, the main simplification made
in these simulations was the use of a tabulated equation of state to follow the thermal evolution
of the gas. The results of the \citet{om05} one-zone model were used to derive internal energy
densities and thermal pressures for the gas at a range of different densities, and this data was
then used to construct a look-up table that could be used by the SPH code to compute the
internal energy and pressure corresponding to a given gas density.

Clark et~al.\ followed the collapse of the gas in their simulation down to a physical scale
of less than an AU (corresponding to a gas density of over $10^{16} \: {\rm cm^{-3}}$). 
Regions collapsing to even smaller scales were replaced by sink particles, created using
the standard \citet{bbp95} prescription. Clark et al.\ showed that at the point at which the
first sink particle formed, the radial profiles of quantities such as the gas density or the
specific angular momentum were very similar to those found in previous studies of
Pop.\ III star formation that were initialized on cosmological scales \citep[e.g.][]{abn02,yoha06}.
They noted that at this point in the simulation, there is no sign of any fragmentation occurring,
and argued that if the simulation were stopped at this point (as would be necessary if the sink particle
technique were not being used), one would probably conclude that the gas would not fragment,
but would merely be accreted by the protostar. However, they show that this is not what actually
happens when the simulation is continued. Instead, the gas fragments, forming 25
separate protostars after only a few hundred years. Clark et al.\ stopped their simulations
after $19 \: {\rm M_{\odot}}$ of gas had been incorporated into sink particles, and showed that
at this point the protostars have masses ranging from $0.02 \: {\rm M_{\odot}}$ to $5 \:
{\rm M_{\odot}}$, but that the mass distribution is relatively flat, with most of the mass locked
up in the few most massive protostars. There is no significant difference between the mass
function of sinks in the low and high resolution calculations, suggesting that fragmentation
is well-resolved in both cases.

This is an intriguing result, but several reasonable concerns could be raised regarding the
numerical technique adopted by \citet{cgk08}. First, the initial conditions for the gas are
an idealized version of what one would find within a real star-forming minihalo, and although
there are indications that the gas loses its memory of the initial conditions prior to fragmentation
occurring, inevitably a few doubts remain. Second, and more importantly, the use of a tabulated
equation of state represents a major simplification of the thermal evolution of the gas, and one
which may make fragmentation more likely to occur. For example, this technique does not allow
one to model the formation of the hot, shocked regions noted by \citet{tna10} in which much or
all of the H$_{2}$ is dissociated, and it is likely to underestimate the temperature of gas falling
in at later times, when the typical infall velocity is larger than during the initial assembly of the
protostar. The Clark et al.\ calculation also neglects the effects of radiative feedback from the
protostars, and assumes that protostars do not merge, even if they come within sub-AU distances
of each other. 

In a follow-up study, \citet{clark11a} addressed one of these concerns -- the use of a tabulated
equation of state -- by performing simulations that replaced this with a detailed treatment of
the chemistry and cooling of primordial gas. In their study, they investigated 
the role that low Mach number turbulence might play in triggering fragmentation in the gas
by performing a set of simulations of the collapse of unstable Bonnor-Ebert spheres. They considered
three initial configurations: a $1000 \: {\rm M_{\odot}}$ cloud with an initial temperature of 300~K;
a $150 \: {\rm M_{\odot}}$ cloud with an initial temperature of 75~K; and a $1000 \: {\rm M_{\odot}}$ 
cloud with an initial temperature of 75~K. In each case, the central density of the Bonnor-Ebert sphere
was taken to be $n_{\rm c} = 10^{5} \: {\rm cm^{-3}}$. The first set of initial conditions were intended
to correspond to the conditions that one would expect to find within one of the first star-forming minihalos, 
while the second set were intended to correspond to the conditions within a minihalo dominated by
HD cooling. Simulations with the third set of initial conditions were run to allow the effects of lowering
the temperature and lowering the total mass to be distinguished. Within the Bonnor-Ebert spheres,
a turbulent velocity field was imposed, with a three-dimensional RMS velocity $\Delta v_{\rm turb}$.

\citet{clark11a} did not claim that this was a completely accurate model of the physical state of the
gas within a real star-forming minihalo. Instead, they treated this study as a kind of physics experiment,
allowing them to investigate the effect of varying a single important parameter -- the turbulent kinetic
energy -- without varying any of the other parameters in the problem, something that it would not be
possible to do if using initial conditions derived from a cosmological simulation. For the first two setups
described above, they performed four simulations, with $\Delta v_{\rm turb} = 0.1, 0.2, 0.4$ and $0.8
\: c_{\rm s}$, respectively, where $c_{\rm s}$ was the initial sound speed. For the third setup (the large,
low temperature clouds), they performed only two simulations, with  $\Delta v_{\rm turb} = 0.4$ and
$0.8 \: c_{\rm s}$, respectively.  The clouds were modelled
using two million SPH particles in each case, yielding a mass resolution of $0.05 \: {\rm M_{\odot}}$
for the $1000 \: {\rm M_{\odot}}$ clouds and $0.0075 \: {\rm M_{\odot}}$ for the $150 \: {\rm M_{\odot}}$ 
clouds. Sink particles were created once the gas density exceeded $10^{13} \: {\rm cm^{-3}}$, and
the sink accretion radius was 20~AU.

\citet{clark11a} found that fragmentation occurred in almost all of their simulated clouds. In the case
of the massive, warm clouds, the only case in which fragmentation did not occur was the simulation
with $\Delta v_{\rm turb} = 0.1 \: c_{\rm s}$. In this simulation, the gas simply collapsed to form a 
single, massive protostar. In the simulations with larger turbulent energies, however,  the
formation of the first protostar was followed within a couple of hundred years by the fragmentation
of the infalling gas and the formation of a significantly larger number of protostars. The relationship
between the turbulent energy and the degree of fragmentation is not straightforward:  the
$\Delta v_{\rm turb} = 0.4 \: c_{\rm s}$ run fragmented more than the $\Delta v_{\rm turb} = 0.2 \: c_{\rm s}$,
as one might expect, but the $\Delta v_{\rm turb} = 0.8 \: c_{\rm s}$ run fragmented {\em less} than
the $\Delta v_{\rm turb} = 0.4 \: c_{\rm s}$ run (although still more than the  $\Delta v_{\rm turb} = 0.2 \: c_{\rm s}$
run). Clark et~al.\ hypothesize that this difference in behaviour is due to the amount of angular 
momentum retained within the collapsing region, which in this case was larger in the 
$\Delta v_{\rm turb} = 0.4 \: c_{\rm s}$ run than in the other runs, but note that this may not always
be the case, and that a much larger series of realizations of the turbulent velocity field would be
needed to make a definitive statement about the relationship between the turbulent energy and the
degree of fragmentation \citep[c.f.][who come to a similar conclusion regarding present-day star
formation]{good04}. The total amount of mass accreted by the sinks is very similar in all four runs, 
and hence the runs that fragment more tend to form lower mass objects than the runs that fragment less. 

In the low-mass, colder clouds, Clark et al.\ find a much lower degree of fragmentation, despite the
fact that the initial number of Jeans masses in these clouds is the same as in the $1000 \:  \index{Jeans mass}
{\rm M_{\odot}}$, $T = 300 \: {\rm K}$ clouds. In this case, fragmentation occurs only in the
$\Delta v_{\rm turb} = 0.2 \: c_{\rm s}$ and $\Delta v_{\rm turb} = 0.8 \: c_{\rm s}$ simulations,
and only a small number of fragments are formed in each case. Clark et al.\ investigate whether
this is due to the lower cloud mass by modelling the collapse of $1000 \: {\rm M_{\odot}}$ clouds
with the same lower initial temperature, and find that although more fragmentation occurs in 
this case, the gas still fragments less than in the  $1000 \: {\rm M_{\odot}}$, $T = 300 \: {\rm K}$ 
case. They suggest that this somewhat counterintuitive behaviour is due to the greater stiffness 
of the effective equation of state in the colder clouds. In both cases, the gas must heat up from
its initial temperature at $10^{5} \: {\rm cm^{-3}}$ to a temperature of roughly 1000~K at 
$10^{10} \: {\rm cm^{-3}}$, and so when the initial temperature is lower, the gas must heat
up more rapidly with increasing density, meaning that it has a larger effective adiabatic index. 
This makes it more difficult to generate the small-scale non-linear structures that are the seeds
for later fragmentation, and also delays the collapse, allowing more of the turbulent energy to 
dissipate. A similar effect has previously been noted by \citet{yoh07} and \citet{to08}, and calls 
into question the common wisdom that minihalos in which the cooling becomes HD-dominated
will inevitably form lower mass stars.

\subsubsection{Models using cosmological initial conditions}
\index{accretion disk|(}
In order to establish whether the fragmentation seen in the \citet{cgk08} model was simply a 
consequence of the highly idealized initial conditions used in that study, several recent follow-up
studies have re-examined the issue using simulations initialized on cosmological scales (i.e.\
scales significant larger than the virial radius of the minihalo). One of the first of these studies 
was carried out by \citet{sgb10}. They first performed a medium resolution cosmological simulation, 
which allowed them to determine the formation site of the first minihalo large enough to cool
effectively and form stars. They then used a hierarchical zoom-in procedure \citep{nw94,tbw97,gao05}
to improve the resolution within a region centered on this formation site, allowing them to achieve a
mass resolution of $1.5 \: {\rm M_{\odot}}$ within the centre of the star-forming minihalo.\footnote{
The value quoted here for the mass resolution of the \citet{sgb10} simulation assumes that
100 or more SPH particles are required to resolve gravitationally bound structures, which is the typical
resolution limit adopted in studies of present-day star formation.  \citet{sgb10} assume that  only 48 SPH 
particles are required, and hence quote a mass resolution that is roughly a factor of two smaller.}
In contrast to \citet{cgk08}, the thermal and chemical evolution of the gas was followed in detail during
the collapse. Once the gravitationally collapsing gas reached a density of $10^{12} \: {\rm cm^{-3}}$, 
it was converted into a sink particle, along with all of the gas within an accretion radius 
$r_{\rm acc} = 50 \: {\rm AU}$. Stacy et al.\ show that following the formation of this first sink, the infalling
gas collapses into a flattened disk. At a time $t = 250 \: {\rm yr}$ after the formation
of the first sink particle, this disk has a radius of 200~AU, but it grows with time and has reached a radius
of 2000~AU by $t = 5000 \: {\rm yr}$, the end of the simulation. H$_{2}$ cooling allows the gas within the 
disk to remain at a temperature of roughly 1000~K, and the disk soon becomes gravitationally unstable,
forming a second sink particle after roughly 300~yr, and a further three sinks after 4000-5000~years
of evolution. At the end of the simulation, the first two sinks to form have become very massive, with
masses of $43 \: {\rm M_{\odot}}$ and $13 \: {\rm M_{\odot}}$ respectively, while the three newer 
sinks still have masses $\sim 1 \: {\rm M_{\odot}}$, close to the resolution limit of the simulation.
\citet{sgb10} do not include the effects of accretion luminosity in their simulation directly, but do assess
its effects during a post-processing stage. They investigate the possible effects of radiation pressure,
but show that this remains unimportant within their simulation throughout the period that they simulate,
in agreement with our analysis above.

The main drawback of the \citet{sgb10} study is their choice of mass resolution. At the point at which it
fragments, the protostellar accretion disk in their simulation has a mass of roughly $35 \: {\rm M_{\odot}}$,
and hence is resolved with only a few thousand SPH particles. This is two orders of magnitude smaller
than the number of particles typically used to model gravitationally unstable accretion disks in the 
context of present-day star formation \citep[see e.g.][]{rla05}, and it is questionable whether a few thousand
particles is enough to properly model the dynamics of the disk. It is therefore possible that the results 
of \citet{sgb10} may have suffered from some degree of artificial fragmentation.

More recently, a study by \citet{clark11b} has dramatically improved the mass resolution used to model
the build-up of a protostellar accretion disk around the first Population III protostar. Clark et al.\ use a
similar basic strategy to Stacy et~al., starting with a medium resolution cosmological simulation to identify
the first star-forming minihalo, and then using a hierarchical zoom-in procedure to improve the resolution
within the gas forming this minihalo. They run this zoomed-in simulation until the maximum density of the
gravitationally collapsing gas reaches $10^{6} \: {\rm cm^{-3}}$. At this point, they extract a spherical region
containing $1000 \: {\rm M_{\odot}}$ of gas from the centre of the minihalo, and resimulate this region at
much higher resolution, using several nested levels of particle splitting \citep{kw02,bl03}. At the final
level of splitting, the particle mass is $10^{-5} \: {\rm M_{\odot}}$ and the mass resolution is 
$10^{-3} \: {\rm M_{\odot}}$, several orders of magnitude better than in the \citet{sgb10} simulation.

\index{feedback!radiative|(}
In addition to the extremely high mass resolution, the other main improvement in the Clark et~al.\ study
compared to previous work is its inclusion of the effects of accretion luminosity
feedback directly within the simulation. To model this, the authors start by writing the bolometric
accretion luminosity produced by a given protostar as
\begin{equation}
L_{\rm acc} = \frac{G \dot{M} M_{*}}{R_{*}},
\end{equation}
where $\dot{M}$ is the accretion rate onto that protostar, $M_{*}$ is the protostellar mass, and $R_{*}$
is the protostellar radius. As Clark et~al.\ attempt to model only the first few hundred years of the
evolution of the gas after the formation of the first protostar, i.e.\ a timescale much less than the protostellar
Kelvin-Helmholtz relaxation timescale, they assume that the protostars remain in the adiabatic accretion
phase of their evolution, with masses and radii that are related by \citep{sps86a}
\begin{equation}
R_{*} = 26 R_{\odot} \left( \frac{M_{*}}{{\rm M}_{\odot}} \right)^{0.27} 
\left(\frac{\dot{M}}{10^{-3} \: {\rm M_{\odot} \: yr^{-1}}} \right)^{0.41}.
\end{equation}
The only remaining uncertainty is then $\dot{M}$, which can be directly measured within the simulation.
Clark et al.\ next assume that the gas is heated by the accretion luminosity at a rate
\begin{equation}
\Gamma_{*} = \rho \kappa_{\rm P} \frac{L_{\rm acc}}{4 \pi r^{2}},
\end{equation}
where $\rho$ is the mass density, $r$ is the distance to the protostar, and $\kappa_{\rm P}$ is the Planck mean 
opacity of the gas, calculated using the tabulated values given in \citet{md05b}. This expression assumes that
the gas is optically thin, and hence will tend to overestimate the heating rate.
\index{feedback!radiative|)}

\begin{figure}
\includegraphics[scale=.60]{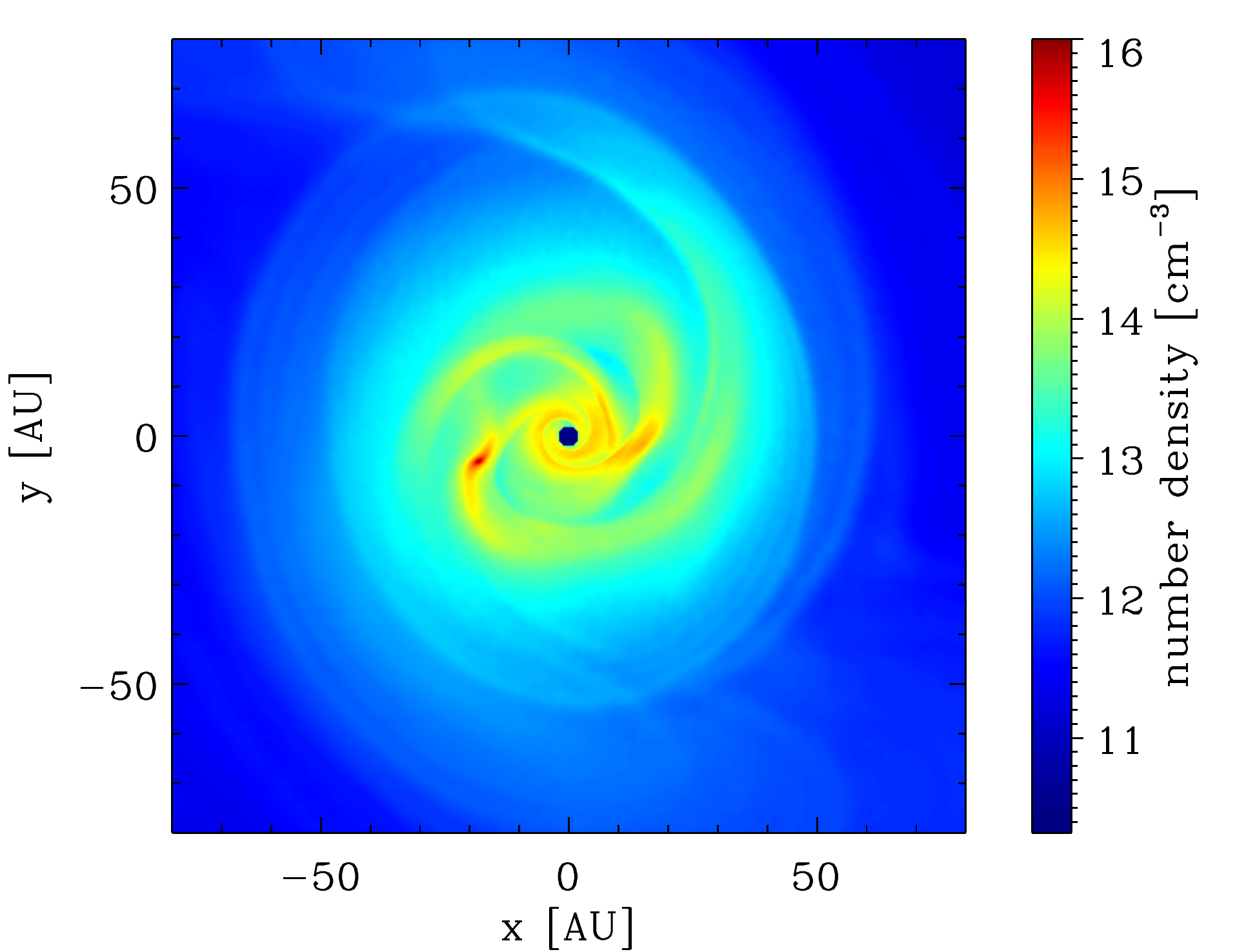}
\caption{The volume density of the protostellar accretion disk in the \citet{clark11a} simulation immediately prior to 
fragmentation. The `hole' in the centre of the disk corresponds to the location of the sink particle representing
the first protostar to form, and occurs because we have not accounted for the mass in the sink particle when
calculating the density. The accretion disk is gravitationally unstable, and has formed several spiral
arms, one of which has begun to fragment.}
\label{fig:disk}  
\end{figure}

\index{sink particles}
Clark et al.\ model protostar formation using sink particles, which are created using the standard \citet{bbp95}
algorithm, with a density threshold $n_{\rm th} = 10^{17} \: {\rm cm^{-3}}$. The sink accretion radius was set to
1.5~AU.  At the point at which the first sink particle forms, the state of the gas in the Clark et al.\ simulation (e.g.\
the density profile and the distribution of specific angular momentum) is very similar to that seen in other high 
resolution simulations of Pop.\  III star formation. However, the authors show that at later times, a protostellar
accretion disk begins to build up around the central protostar, just as in the Stacy et al.\ study. The significantly
higher resolution of the Clark et al.\ simulation allows them to model the build up of this disk on scales much
closer to the central protostar, and to resolve the disk with a far larger number of SPH particles. The growth of
the accretion disk is followed for around 100~years after the formation of the first protostar, and Clark et al.\ show
that after around 90~years (corresponding to around 1.5 orbital periods for the disk), the accretion disk begins to 
fragment, forming several low-mass protostars. At the time at which it fragments, the disk contains several solar 
masses of gas (and hence is resolved with several hundred thousand SPH particles), compared to around $0.4 \:
{\rm M_{\odot}}$ in the central protostar, and the disk radius is a few tens of AU. The state of the disk at the
onset of fragmentation is illustrated in Figure~\ref{fig:disk}.

Clark et~al.\ argue that the reason that the protostellar accretion disk fragments is that it is unable to transfer 
gas onto the protostar fast enough to keep up with the rate at which fresh gas is falling onto the disk. 
This causes the surface density of the disk to increase, which eventually results in it becoming gravitationally 
unstable and fragmenting. This argument can be made somewhat more quantitative if one treats the disk using
the standard \citet{ss73} thin disk model, and hence writes the mass flow rate through the disk at a radius $r$ as
\begin{equation}
\dot{M}(r) = 3 \pi \alpha c_{\rm s}(r) \Sigma(r) H(r),
\end{equation}
where $\alpha$ is the viscosity parameter, and $c_{\rm s}(r)$, $H(r)$ and $\Sigma(r)$ are the sound speed, 
disk thickness and surface density, respectively, at a radius $r$. Clark et al.\ use the values of $c_{\rm s}$, $H$ 
and $\Sigma$ 
provided by their simulation to show that $\dot{M}(r)$ is smaller than the accretion rate onto the disk for a wide
range of radii, even if one adopts $\alpha = 1$ (which already presupposes that the disk is gravitationally 
unstable). The growth of the disk therefore appears to be unavoidable, leading to its fragmentation once portions
of it develop a Toomre stability parameter $Q < 1$, where $Q \equiv c_{\rm s} \kappa / \pi G \Sigma$, with
$\kappa$ here being the epicyclic frequency.

Previous semi-analytical studies of the structure of Pop.\ III accretion disks came to a somewhat 
different \index{molecular hydrogen!cooling|(}
conclusion regarding their stability, predicting that $Q \gg 1$ throughout the disk \citep{tm04,tb04,md05}.
However, these models neglected the effect of H$_{2}$ cooling, motivated by the assumption that the 
H$_{2}$ content of a Pop.\ III protostellar accretion disk is negligible (J.\ Tan, private communication). 
This assumption leaves H$^{-}$ as the primary source of opacity at temperatures $T \sim 7000 \: {\rm K}$ 
and below \citep{lcs91,md05b}. For gas in chemical equilibrium, the opacity of H$^{-}$ decreases sharply
with decreasing temperature, and consequently these models find the equilibrium temperature
of the accretion disk to be high, $T \sim 6000 \: {\rm K}$. In comparison, Clark et~al.\ show that when 
the effects of H$_{2}$ are taken into account, the characteristic temperature of the gas in the disk lies in
the range of 1500 to 2000~K. The disks in these previous semi-analytical models could therefore 
transfer mass onto the protostar more rapidly than the disk in the Clark et~al.\ simulation, owing to their 
higher sound-speed and larger thickness, but at the same time were also more stable against gravitational
fragmentation. It is therefore not surprising that these previous studies predicted that the accretion disk
should be stable, and highlights the crucial role played by H$_{2}$ cooling in enabling disk fragmentation.
\index{molecular hydrogen!cooling|)}

\index{feedback!radiative|(}
In addition to the simulation described above, in which the value of $\dot{M}$ used to calculate the
accretion luminosity was measured directly, Clark et al.\ also performed two additional simulations in 
which $\dot{M}$ was kept fixed, allowing them to investigate the role played by accretion luminosity
heating. They considered cases with $\dot{M} = 10^{-3} \: {\rm M_{\odot}} \: {\rm yr^{-1}}$ (somewhat
smaller than the measured value) and  $\dot{M} = 10^{-2} \: {\rm M_{\odot}} \: {\rm yr^{-1}}$ (larger
than the measured value), and showed that as $\dot{M}$ (and hence $L_{\rm acc}$) increase, the
disk becomes warmer and thicker and takes longer to fragment. However, fragmentation still occurs
in every case, demonstrating that accretion luminosity heating is unable to prevent the disk from
fragmenting \citep[c.f.][who argues that it plays a crucial role in suppressing fragmentation in local
star-forming systems]{krum06}.
\index{feedback!radiative|)}
 
One drawback of the \citet{clark11b} study is that they examined only a single star-forming minihalo,
and although they showed that the properties of this halo (e.g.\ mass, spin parameter, formation redshift)
were similar to those of the minihalos modelled in previous studies of Pop.\ III star formation, nevertheless
the suspicion remains that perhaps this particular minihalo was unusual in some way. This concern was
addressed by \citet{greif11a}. They used the new moving-mesh code AREPO \citep{spring10} to
study Pop.\ III star formation in five different minihalos, using a sink particle algorithm to allow them to
follow the evolution of the gas past the point at which the first protostar formed. In all five of the systems
that they modelled, they found similar behaviour to that in the Clark et~al.\ study: an accretion disk built
up around the first protostar, became gravitationally unstable, and began to fragment after only a few years.
These results suggest that Clark et~al.\ were right to claim that disk fragmentation, and the
resulting formation of Pop.\ III binary systems, or higher order multiple systems, is an almost inevitable
outcome of Population III star formation. 

\index{sink particles|(}
\citet{greif11a} also examined the issue of whether the objects represented by the individual sink particles
would truly survive as separate protostars, or whether they would simply merge into a single massive
protostar as the system evolved further. They considered two different schemes for merging sink particles.
In the  standard scheme, sink particles coming within a distance of $100 \: {\rm R_{\odot}}$ of each other 
were merged to form a single sink, provided that the total energy of the two-body system was negative.
In an alternative model, utilizing what Greif et~al.\ dub as ``adhesive'' sinks, the energy check was omitted,
and sinks were always merged when within a distance of $100 \: {\rm R_{\odot}}$ of each other. Greif
et~al.\ justify their choice of this critical distance in two ways: first, it is also the accretion radius adopted
for their sinks, meaning that the gas flow on smaller scales close to the sinks is not resolved; and second,
it is roughly equal to the maximum size of a pre-main sequence protostar predicted by the models discussed
in Section~\ref{proto}.
\index{sink particles|)}

The majority of the protostars formed in the Greif et~al.\ simulations have at least one close encounter
with another protostar, but when the standard merger algorithm is used, many of these encounters 
result in a purely dynamical interaction, as the total energy of the protostellar pair is too large to allow
them to merge. On the other hand, when the adhesive sinks are used, many of these encounters lead
to mergers. Greif et al.\ show that although the total mass incorporated into protostars is roughly the 
same in both cases, the number of protostars that survive as individual objects is reduced by a factor
of up to a few, and the mean protostellar mass is consequently higher. In both cases, the protostars
have a relatively flat mass distribution, with most of the protostellar mass  being accounted for by a
small number of high-mass protostars. The protostars have a broad distribution of radial velocities,
ranging from $v_{\rm rad} \sim 1 \: {\rm km} \: {\rm s^{-1}}$ to $v_{\rm rad} \sim 100  \: {\rm km} \: 
{\rm s^{-1}}$, and in many cases the radial velocity is greater than the escape velocity of the central
region of the minihalo. It is likely that this leads to a significant fraction of the Pop.\ III protostars escaping
from the minihalo entirely, although Greif et~al.\ do not follow their evolution for long enough to confirm
this. It is possible that these protostars will accrete very little additional gas once they escape from the 
high density region at the centre of the minihalo \citep[see e.g.][]{jk11}, in which case
their final masses would be very similar to the masses that they have at the point at which they are ejected. 
Greif et~al.\ show that in the standard case, a considerable
number of these ejected protostars have masses $M < 1 \: {\rm M_{\odot}}$. If these protostars do indeed
avoid accreting further gas after their ejection, then they would have
lifetimes that are comparable to the current age of the Universe. This suggests that it
may be possible for some Population III stars to survive until the present day. However, when the
adhesive sinks are used, the number of ejected protostars with subsolar masses is greatly reduced,
demonstrating that this conclusion is highly sensitive to our treatment of protostellar mergers.

\index{feedback!radiative|(}
The majority of the Greif et~al.\ simulations did not include the effects of the accretion luminosity 
generated by the collection of protostars. However, they did consider one case in which this was
included, using the Clark et~al.\ treatment with a fixed value for the accretion rate used in the
determination of the accretion luminosity, $\dot{M} = 0.1 \: {\rm M_{\odot}} \: {\rm yr^{-1}}$. The
effect of this was to puff up the disk, causing fragmentation to occur at a slightly larger distance
from the initial protostar. However, despite the unrealistically high value adopted for $\dot{M}$,
fragmentation still occurred and the number of protostars that formed was barely affected.

A more detailed study of the effects of accretion luminosity heating was carried out by \citet{smith11}. 
They used Gadget to resimulate the central 2~pc of the
minihalos simulated by \citet{greif11a}, starting at a time prior to protostar formation at which the
peak density of the gas was around $10^{9} \: {\rm cm^{-3}}$. \citet{smith11} evolved these systems past
the time at which the first protostar formed, and used sink particles with large accretion radii 
($r_{\rm acc} = 20 \: {\rm AU}$) to allow them to follow the dynamical evolution of the system for
an extended period. For each of the five minihalos, they performed simulations both with and
without accretion luminosity heating. When the accretion luminosity heating was included, it was
treated in the same fashion as in \citet{clark11b}. They found that in general, the effect of the
accretion luminosity heating was to delay fragmentation and reduce the number of fragments
formed. However, they also showed that the effect was relatively small, and had less influence
on the number of fragments formed than did the intrinsic variation in halo properties arising 
from their different assembly histories. 
\index{feedback!radiative|)}
\index{accretion disk|)}

\subsubsection{Open questions}
\label{open}
As the discussion in the previous section has shown, the past couple of years has seen a large
increase in the number of simulations of Pop.\ III star formation that show evidence for fragmentation,
suggesting that the older picture that had Pop.\ III stars forming in isolation with masses of 
$100 \: {\rm M_{\odot}}$ or more is in need of some revision. However, many aspects of the
fragmentation scenario remain unclear. Some of the most important open questions are summarized
below.
\begin{itemize}
\item Are our treatments of optically thick H$_{2}$ cooling and accretion luminosity heating adequate?
\end{itemize}
\index{molecular hydrogen!cooling|(}
\index{feedback!radiative|(}
The fragmentation of the gas observed in these simulations typically occurs at densities at which 
H$_{2}$ line cooling is optically thick, and hence may depend on the method used to account for
the reduction that this causes in the cooling rate. At present, both of the methods in common usage
represent relatively crude approximations, and it remains to be seen whether the behaviour of the
gas will remain the same if a more accurate treatment is used. Similarly, the method currently used
to treat the effects of accretion luminosity heating also makes a number of major simplifications
that may influence the outcome of the simulations.
\index{molecular hydrogen!cooling|)}
\index{feedback!radiative|)}
\begin{itemize}
\item What role do magnetic fields play?
\end{itemize}
\index{magnetic fields|(}
If a non-negligible magnetic field can be generated by the turbulent dynamo during the gravitational
collapse of the gas, as discussed in Section~\ref{magn}, then this may influence the evolution and stability of
the disk. The presence of a magnetic field may make the disk unstable via the magnetorotational
instability \citep{tb04,sl06}, although the resulting mass transfer onto the protostar is unlikely to be fast 
enough to prevent the disk from becoming gravitationally unstable. A more important effect may be
magnetic braking of the infalling gas, which could act to significantly reduce the angular momentum
of the gas reaching the disk \citep[see e.g.][]{hc09}. Whether either of these effects can significantly suppress
fragmentation remains to be determined.
\index{magnetic fields|)}
\begin{itemize}
\item How often do Population III protostars merge? 
\end{itemize}
Current simulations either ignore mergers entirely \citep[e.g.][]{clark11b,smith11}, or treat them using very
simple approximations that do not properly account for the effects of tidal forces
\citep[e.g.][]{greif11a}. However, it is clear from the results of the Greif et~al.\ study
that the method used to treat mergers has a significant influence on the number of protostars that survive, 
their mass distribution and their kinematics. Improving the accuracy with which protostellar mergers are
treated within this kind of simulation is therefore an important priority.
\begin{itemize}
\item Can we find some way to do without sink particles?
\end{itemize}
\index{sink particles|(}
The concerns outlined above regarding the way in which protostellar mergers are treated would be
greatly ameliorated if we were able to run the simulations without sink particles, as in this case we
would be able to model directly how the gas behaves on scales of the order of $100 \: {\rm R_{\odot}}$.
To do this, however, it will be necessary to devise some scheme for treating these very small scales
that does not fall foul of the Courant time constraint discussed previously. \citet{greif12} have recently 
published the results of an initial effort along these lines, but were only able to simulate the first ten
years of the evolution of the disk, owing the extremely high computational cost of the calculation.

\index{sink particles|)}
\begin{itemize}
\item How rapidly does H$_{2}$ photodissociation occur?
\end{itemize}
\index{molecular hydrogen!chemistry|(}
\index{feedback!radiative|(}
\index{photodissociation|(}
Fragmentation is dependent on the cooling provided by H$_{2}$ and does not occur in models of protostellar
accretion disks that omit this effect \citep{tm04,md05}. It is therefore highly probable that fragmentation will
cease once the H$_{2}$ has been photodissociated by Lyman-Werner band photons emitted from any massive
protostars that form. What is not yet clear is how quickly this will occur.  \citet{mt08} show that the number of 
Lyman-Werner band photons produced by a zero-age main sequence Population III star increases sharply
with increasing stellar mass, before levelling off at a value $S_{\rm lw} \sim 10^{49} \: {\rm photons \: s^{-1}}$
for $M_{*} \sim 30 \: {\rm M_{\odot}}$. If we assume that all of these photons are absorbed by H$_{2}$ and
that 20\% of these absorptions lead to dissociation \citep{db96}, then the radiation from the star will 
photodissociate H$_{2}$ at a rate $\dot{M}_{\rm dis} = 0.1 \: {\rm M_{\odot}} \: {\rm yr^{-1}}$, leading to
complete removal of the H$_{2}$ within only a few hundred years. However, it is likely that many of the
available photons will not be absorbed by H$_{2}$, either because they never coincide with one of the
Lyman-Werner band lines, or because they escape along a direction in which most of the H$_{2}$ has
already been dissociated, or because they are absorbed by atomic hydrogen \citep{gb01}, and so the
dissociation time for the H$_{2}$ could be significantly longer than suggested by this simple estimate,
particularly once one accounts for the effects of three-body H$_{2}$ formation.
\index{photodissociation|)}
\index{molecular hydrogen!chemistry|)}
\begin{itemize}
\item How rapidly is the gas ionized? Does this completely suppress accretion, or simply suppress
fragmentation?
\end{itemize}
\index{photoionization|(}
As we have already discussed in Section~\ref{feed}, the most plausible mechanism for shutting off the supply 
of cold gas at the centre of the minihalo is the formation of an HII region whose thermal pressure is
sufficient to expel most of the gas. However, only a few studies have looked at the 
interaction between the growth of the HII region and the evolution of the
protostellar accretion disk. In particular, the issue has not yet been looked at within the context of the
fragmentation model discussed above. It is therefore unclear how rapidly the HII region will grow,
and whether it will immediately act to shut off accretion, or whether pockets of dense, cold gas can
survive within the HII region for an extended period.
\index{photoionization|)}
\index{feedback!radiative|)}
\begin{itemize}
\item Do any low-mass Population III stars survive until the present day?
\end{itemize}
One of the most exciting results of the \citet{greif11a} model is that some of the protostars that are ejected
from the centre of the star-forming minihalo have masses that are below $0.8 \: {\rm M_{\odot}}$, and hence
lifetimes that are longer than the current age of the Universe. If these protostars avoided accreting any further
gas, then they could have survived until the present-day, raising the possibility of directly detecting truly
metal-free stars within the Milky Way. However, as we have already discussed above, the number of
protostars with sub-solar masses that are ejected from the star-forming region is very sensitive to the way in
which protostellar mergers are treated, and hence is highly uncertain at present. In addition, it is possible
that any Pop.\ III protostars that have survived until the present day have also become too polluted with
metals by ongoing accretion from the ISM for us to recognize them as metal-free stars, although the best 
current estimates \citep{fjb09,jk11} suggest that the effects of pollution will be very small.
\index{fragmentation|)}

\section{Summary}
In this review, we have focussed on three main topics: how the first star-forming minihalos come into
existence and why they have the properties that they do; how gas within a representative minihalo
cools, collapses and forms a protostar; and how this protostar and the massive clump of gas
surrounding it subsequently evolve. 

We have seen that on large scales, we now have a relatively 
good understanding of the physical processes involved in the formation of the first star-forming 
minihalos. In order for gas to accumulate within a dark matter minihalo, it must be able to overcome
the effects of both gas pressure and also the large-scale streaming motion of the gas relative to the 
dark matter. Since this streaming motion is typically supersonic, the latter effect generally dominates,
and the result is that gas is prevented from accumulating in large quantities within minihalos with
masses of less than around $10^{5} \: {\rm M_{\odot}}$. Within more massive minihalos, the gravitational
force exerted by the dark matter is strong enough to overcome the effects of the gas pressure and
the coherent streaming, and the gas begins to undergo gravitational collapse, reaching densities
that are several hundred times higher than the cosmological background density. As the gas
collapses, however, it is heated by compression and shocks. In order for the collapse to continue,
the gas must be able to dissipate this energy, which it does through rotational and vibrational
line emission from H$_{2}$.

\index{molecular hydrogen!cooling|(}
\index{molecular hydrogen!chemistry|(}
In Section~\ref{coolchem}, we saw that the amount of H$_{2}$ formed within a given minihalo is a 
strong function of the temperature of the gas, with the final molecular fraction scaling roughly as $x_{\rm H_{2}}
\propto T^{3/2}$ with the temperature $T$. The H$_{2}$ cooling rate is also a strong function of
temperature. As a result, one finds that there is a critical minihalo virial temperature, $T_{\rm crit} \sim 1000
\: {\rm K}$ marking the division between cooler halos that do not dissipate much energy within a Hubble
time, and hence which do not form stars, and warmer halos that do manage to cool and form stars. 
As the virial temperature of a minihalo is a simple function of its mass and redshift, one can derive a
critical minihalo mass that must be exceeded in order for the gas to cool effectively. This critical mass
scales approximately as $M_{\rm crit} \sim 1.6 \times 10^{6} (1+z / 10)^{-3/2} \: {\rm M_{\odot}}$,
given standard values for the cosmological parameters. Combining this constraint with that arising 
from coherent streaming, one finds that at redshifts $z > 40$, the minimum mass of a star-forming 
minihalo is set by the need to overcome the effects of the streaming, and is roughly $10^{5} \: {\rm M_{\odot}}$, 
while at $z < 40$, H$_{2}$ cooling is the limiting factor, and the minimum mass scale is somewhat larger.

On smaller scales, we have also developed an increasingly good understanding of how the gas evolves as 
it cools, undergoes runaway gravitational collapse, and forms the first protostar. As outlined in Section~\ref{collapse},
the gas first passes through a ``loitering'' phase, during which cold gas accumulates at the centre of the
minihalo. The temperature and density of the gas at this point  depend on the nature of the dominant
coolant. When H$_{2}$ dominates, we have $T \sim 200 \: {\rm K}$ and $n \sim 10^{4} \: {\rm cm^{-3}}$,
while if HD dominates, then $T \sim 100 \: {\rm K}$ and $n \sim 10^{6} \: {\rm cm^{-3}}$. The
loitering phase ends and the collapse of the gas accelerates once the mass of cold gas that has accumulated 
exceeds the local value of the Bonnor-Ebert  mass, which is around $1000 \: {\rm M_{\odot}}$ in the 
H$_{2}$-dominated case, but only $40 \: {\rm M_{\odot}}$ in the HD-dominated case. The next major event
to occur is the onset of three-body H$_{2}$ formation at $n \sim 10^{8} \: {\rm cm^{-3}}$
which rapidly converts most of the atomic hydrogen into H$_{2}$. The associated heat input leads to an increase 
in the gas temperature to $T \sim 1000$-2000~K, with the details depending to a significant extent on the 
rate coefficient chosen for reaction~\ref{3b1}, which is poorly constrained at low temperatures. At $n \sim 10^{10}
\: {\rm cm^{-3}}$, the gas becomes optically thick in the main H$_{2}$ cooling lines, but remains optically
thin in the continuum. It can therefore continue to cool reasonably effectively at these densities, with the
mean temperature only rising relatively slowly with increasing density. At $n \sim 10^{14} \: {\rm cm^{-3}}$,
a new process, collision-induced emission from H$_{2}$, begins to dominate the cooling. However, this
does not lead to a significant drop in the gas temperature, as the gas quickly becomes optically thick in the
continuum. At densities above $n \sim 10^{16} \: {\rm cm^{-3}}$, further radiative cooling of the gas is ineffective
and the only remaining process capable of slowing the temperature rise is collisional dissociation of the
H$_{2}$. While the H$_{2}$ fraction in the gas remains significant, the temperature is prevented from rising
much above 3000~K, but once most of the H$_{2}$ has been destroyed, the temperature in the core rises
steeply, and the internal thermal pressure eventually becomes strong enough to halt the collapse. 
State-of-the-art simulations have followed the gravitational collapse of the gas up to this point, which we
can identify as the moment at which the first true Population III protostar forms.  
\index{molecular hydrogen!cooling|)}
\index{molecular hydrogen!chemistry|)}

Nevertheless, several uncertainties remain in this picture of Pop.\ III star formation. As already noted, the
uncertainty in the three-body H$_{2}$ formation rate limits the accuracy with which we can model the
chemical and thermal evolution of the collapsing gas. In addition, current three-dimensional collapse
models make use of simplified treatments of the effect of opacity on the H$_{2}$ cooling rate, and the
uncertainty that this introduces into the models has not yet been properly quantified. 
Further uncertainty comes from two additional issues which have only recently begun to be addressed:
the role played by magnetic fields, and the influence of dark matter annihilation. Although the strength
of any seed magnetic field existing prior to the assembly of the first star-forming minihalos is still poorly
constrained, it now seems clear that the small-scale turbulent dynamo acting during the collapse of the
gas will rapidly amplify even a very weak initial field up to a point at which it could potentially become
dynamically significant. However, neither the final strength of the field nor its correlation length are well 
constrained at present, and without a better understanding of these values it is difficult to say to what 
extent the magnetic field will influence the details of the collapse. The role played by heating
and ionization due to dark matter annihilation is even less well understood. Simple models suggest
that it may be extremely important and may result in the formation of ``dark stars'' supported by the \index{dark star}
energy released by dark matter annihilation rather than by nuclear fusion, but the only hydrodynamical
study performed to date suggests that the influence on the collapse is small, and that dark stars do not
actually form. 

Finally, there remains the question of how the gas evolves after the formation of the first protostar.
For much of the last decade, the leading model for this phase of the evolution of the gas has been
what we have termed the ``smooth accretion'' model. In this model, it is assumed that the gas
surrounding the newly formed protostar does not fragment, but instead simply smoothly accretes
onto the protostar, primarily via a protostellar accretion disk. Considerable work has been done
within the framework of this model to understand the structure of the protostar during the accretion
phase, and the effect of protostellar feedback on the surrounding gas. This work has shown that  \index{feedback!radiative}
any feedback occurring prior to the protostar joining the main sequence is unlikely to significantly
reduce the accretion rate, and that the most plausible mechanism for terminating the accretion is
photoionization of the accretion disk by ionizing radiation from the central star, implying that it \index{photoionization}
must already have grown to some tens of solar masses. This model therefore predicts that Pop.\
III stars will generally be solitary, with only one or two forming in each minihalo, and massive,
with masses $M \gg 10 \: {\rm M_{\odot}}$.

\index{fragmentation|(}
Over the past couple of years, however, several new studies have appeared that have cast 
considerable doubt on the smooth accretion model. These studies have attempted to directly
model the evolution of the gas as it begins to be accreted, and have shown
that the accretion disk that builds up around the protostar is unstable to gravitational
fragmentation even if the stabilizing effects of accretion luminosity feedback from the central  \index{feedback!radiative}
protostar are taken into account. Once a few fragments have formed, the dynamical interactions
between the individual fragments and between the fragments and the gas can lead to further
fragmentation, and to the ejection of low-mass fragments from the system. If we assume that 
all of the gravitationally bound fragments form protostars, then the result of this model
is the assembly of a small, extremely dense cluster of Pop.\ III protostars with a wide range
of masses. As discussed in Section~\ref{open},  many aspects of the fragmentation scenario remain 
unclear and much work remains to be done before we can hope to have a good understanding
of the final protostellar mass function. Nevertheless, these results suggest that Population III
star formation perhaps has far more in common with present-day star formation than has been
previously recognised.
\index{fragmentation|)}

\section*{Acknowledgments}
The author would like to thank a large number of people with whom he has had interesting and
informative discussions about the physics of Population III star formation, including T.\ Abel, V.\ Bromm, P.\ Clark, 
G.\ Dopcke, T.\ Greif,  Z.\ Haiman, T.\ Hosokawa, R.\ Klessen, M.\ Norman, K.\ Omukai, B.\ O'Shea, D.\ Schleicher, 
B.\ Smith,  R.\ Smith, A.\ Stacy,  J.\ Tan, M.\ Turk, D.\ Whalen, and  N.\ Yoshida. Special thanks also go to R.\ Smith 
and M.\ Turk for providing some of the data plotted in Figure~\ref{fig:accrete}, and to P.\ Clark for his assistance 
with the production of Figure~\ref{fig:disk}. Financial support for this work was provided by the 
Baden-W\"urttemberg-Stiftung via their program International Collaboration II (grant P-LS-SPII/18), from the German 
Bundesministerium f\"ur Bildung und Forschung via the ASTRONET project STAR FORMAT (grant 05A09VHA), and 
by a Frontier grant of Heidelberg University sponsored by the German Excellence Initiative.

%%%%%%%%%%%%%%%%%%%%%%%% referenc.tex %%%%%%%%%%%%%%%%%%%%%%%%%%%%%%
% sample references
% %
% Use this file as a template for your own input.
%
%%%%%%%%%%%%%%%%%%%%%%%% Springer-Verlag %%%%%%%%%%%%%%%%%%%%%%%%%%
%
% BibTeX users please use
% \bibliographystyle{}
% \bibliography{}
%

%\printindex

\end{document}